\documentclass[preprint,aps,prd,showpacs,nofootinbib,superscriptaddress]{revtex4}
\usepackage{graphicx}
\usepackage{amsmath}
\usepackage{amssymb}
\usepackage{bm}
\usepackage{bbm}
\begin{document}
\preprint{ANL-HEP-PR-07-20}
\title{\mbox{}\\[10pt]
Inclusive Charm Production in
$\bm{\chi_{b}}$ Decays
}
\author{Geoffrey T. Bodwin}
\affiliation{
High Energy Physics Division, Argonne National Laboratory,\\
9700 S. Cass Avenue, Argonne, Illinois 60439, USA}

\author{Eric Braaten}
\affiliation{
Physics Department, Ohio State University,
Columbus, Ohio 43210, USA}

\author{Daekyoung Kang}
\affiliation{
Physics Department, Ohio State University,
Columbus, Ohio 43210, USA}
\affiliation{
Department of Physics, Korea University,
Seoul 136-701, Korea}

\author{Jungil Lee}
\affiliation{
High Energy Physics Division, Argonne National Laboratory,\\
9700 S. Cass Avenue, Argonne, Illinois 60439, USA}
\affiliation{
Department of Physics, Korea University,
Seoul 136-701, Korea}

\date{\today}
\begin{abstract}
We calculate the inclusive decay rate of the spin-triplet 
bottomonium states $\chi_{bJ}$ 
into charm hadrons, including the leading-order 
color-singlet and color-octet $b \bar b$ annihilation mechanisms. 
We also calculate the momentum distribution of the charm quark 
from the decay of $\chi_{bJ}$.  
The infrared divergences from the color-singlet process 
$b \bar b \to c \bar c g$ are factored into the probability 
density at the origin for a $b \bar b$ pair in a color-octet state.
That probability density can be determined phenomenologically
from the fraction of decays of $\chi_{bJ}$ that include charm hadrons.
It can then be used to predict the partial widths into light hadrons 
for  all four states in the $P$-wave bottomonium multiplet. 
\end{abstract}


\maketitle

\section{Introduction}

The asymptotic freedom of QCD suggests that the total widths of heavy
quarkonium states should be calculable using perturbation theory.
The earliest calculations of the widths of $P$-wave quarkonium states using
perturbative QCD were plagued with infrared divergences
\cite{Barbieri:1975am,Barbieri:1976fp,Barbieri:1981xz}.
The calculations were based on a factorization assumption that the 
width could be expressed as the product of  $|R'(0)|^2$, where
$R'(0)$ is the derivative of the radial wave function at the origin,
and a perturbatively calculable coefficient.
However the coefficients were found to be 
infrared divergent
at leading order in $\alpha_s$ for the spin-1 states and at 
next-to-leading order in $\alpha_s$ for the spin-0 and spin-2 states.
The infrared divergences were often expressed in terms of a 
logarithmic dependence on the binding energy of the quarkonium, 
a quantity that is not calculable using perturbation theory.  
However the correct interpretation of the infrared divergences 
is that they reveal the failure of the factorization assumption.

This problem was overcome in 1992 when Bodwin, Braaten, and Lepage
showed that the infrared divergences could be absorbed into the
probability for the heavy-quark-antiquark ($Q \bar Q$) pair to be at the
same point in a color-octet state \cite{Bodwin:1992ye}.  
They used a nonrelativistic effective field theory for the $Q \bar Q$ 
sector of QCD called NRQCD to derive a general factorization formula
for inclusive quarkonium decay rates \cite{Bodwin:1994jh}.  
A $P$-wave multiplet consists of four heavy quarkonium states:
$\chi_{Q0}$, $\chi_{Q1}$, $\chi_{Q2}$, and $h_Q$ with $J^{PC}$ quantum
numbers $0^{++}$, $1^{++}$, $2^{++}$, and $1^{+-}$, respectively. At
leading order in the velocity $v$ of the heavy quark or antiquark in the
quarkonium rest frame, there are only two independent nonperturbative
factors in the annihilation decay rates of all four states in the
$P$-wave multiplet: $\langle \mathcal{O}_1\rangle$, which is proportional
to $|R'(0)|^2$, and $\langle \mathcal{O}_8 \rangle$, which is proportional
to the probability for the $Q$ and $\bar Q$ to be at the same point in a
color-octet state.  These nonperturbative factors can be expressed as
matrix elements of local four-quark operators in NRQCD. The short-distance
coefficients of the NRQCD matrix elements can be calculated as power
series in the QCD coupling constant $\alpha_s$.

The widths of all four states in a $P$-wave multiplet can be calculated
by using the NRQCD factorization formula, 
once the two nonperturbative factors
$\langle \mathcal{O}_1 \rangle$ and $\langle \mathcal{O}_8 \rangle$ 
have been determined. These matrix elements can be calculated by using 
lattice simulations of NRQCD.  An alternative 
is to estimate the color-singlet matrix element 
$\langle \mathcal{O}_1  \rangle$ by using potential models 
and to determine the color-octet matrix element 
$\langle \mathcal{O}_8 \rangle$ phenomenologically.
The phenomenological determination of $\langle \mathcal{O}_8 \rangle$ 
requires the measurement of an observable that is sensitive 
to this matrix element.  In the case of bottomonium,
one such observable is the inclusive rate for charm production 
in decays of the spin-triplet $P$-wave states $\chi_{bJ}$. 
This rate is sensitive to $\langle \mathcal{O}_8 \rangle$ 
because the production of charm quarks from $b \bar b$ annihilation 
in the color-singlet channel is suppressed by a
factor of $\alpha_s$, relative to production in the color-octet channel.

There has been little previous work on open charm production in bottomonium
decays.  In 1978, Fritzsch and Streng calculated the decay rate of $\Upsilon$ 
into charm at leading order in $\alpha_s$ \cite{Fritzsch:1978ey}.  
In 1979, Barbieri, Caffo, and Remiddi calculated the decay rates of the
$P$-wave bottomonium states into charm at leading order
in $\alpha_s$ under the assumption that the rates could be expressed 
as products of $|R'(0)|^2$ and a perturbatively calculable coefficient
\cite{Barbieri:1979gg}.
In the case of $\chi_{b0}$ and  $\chi_{b2}$, the coefficients contained
infrared divergences that were expressed in terms of logarithms
of the binding energy.  However, as we have mentioned, the correct 
interpretation of the infrared divergences is that they are contained 
in the probability to find the $Q\bar{Q}$ pair at a point in a 
color-octet state. By making use of the NRQCD factorization formalism,
one can now carry out rigorous calculations of inclusive charm production
from $\chi_{bJ}$ decays.

In their 1979 paper, Barbieri, Caffo, and Remiddi calculated the 
invariant mass distribution of the $c \bar c$ pair in $\chi_{bJ}$ decays.
In order to make contact with experiment, one might be tempted to identify
this distribution with the invariant mass distribution of pairs of 
charm hadrons.  However that distribution cannot be measured easily
because the probability of identifying both charm hadrons is very low.
Furthermore, the effects of the hadronization of the charm quark 
into a charm hadron have a large effect on the distribution.
These effects cannot be calculated perturbatively, and they would also 
be very difficult to measure.  A more useful quantity to calculate is 
the momentum distribution of the charm quark in $\chi_{bJ}$ decays.  
This cannot be compared directly with the momentum distribution of 
the charm hadrons because of the effects of hadronization.  
However, the effects of hadronization can be determined 
experimentally by measuring the momentum distribution of charm hadrons 
in $e^+ e^-$ annihilation.

On the experimental side, the spin-triplet members of two multiplets 
of $P$-wave bottomonium states have been discovered:  $\chi_{bJ}(1P)$ 
and $\chi_{bJ}(2P)$.  The only properties of these states that have 
been measured thus far are their masses and their radiative branching 
fractions into the $S$-wave bottomonium states $\Upsilon(nS)$. 
The total widths of the $\chi_{bJ}(nP)$ states have not been measured.
Recent runs of the CLEO experiment at the $\Upsilon(2S)$ and $\Upsilon(3S)$
resonances have provided new data on the $\chi_{bJ}(1P)$ and 
$\chi_{bJ}(2P)$ states. The $B$-factory experiments BABAR and Belle 
can study the $\chi_{bJ}(nP)$ states by using data samples of 
$\Upsilon(2S)$ and $\Upsilon(3S)$ provided by initial-state radiation.
The Belle experiment has also accumulated data by running directly 
on the $\Upsilon(3S)$ state.

In this paper, we study inclusive charm production in $P$-wave bottomonium 
decays.  In Sec.~\ref{sec:NRQCD}, we present the NRQCD factorization 
formulas for the annihilation decays of $P$-wave bottomonium states,
and we discuss the NRQCD matrix elements that appear as long-distance 
factors in the factorization formulas.
In Sec.~\ref{sec:momentum}, we calculate the charm-quark momentum 
distribution in decays of the spin-triplet $P$-wave states 
$\chi_{bJ}$.  We include the color-singlet process $b \bar b \to c \bar c g$, 
which has a short-distance coefficient of order $\alpha_s^3$, 
and the color-octet process $b \bar b \to c \bar c$, which has a 
short-distance coefficient of order $\alpha_s^2$. 
In Sec.~\ref{sec:rate}, we calculate the inclusive rate into charm 
by integrating over the charm-quark momentum distribution.  
In Sec.~\ref{sec:meson}, we illustrate the momentum distribution 
for a charm meson $D$ by convolving the charm-quark momentum distribution 
with a fragmentation function for $c \to D$ that has been measured 
in $e^+e^-$ annihilation. 
Details of the calculations are presented in appendices.

\section{Annihilation decays of $\bm{P}$-wave bottomonium}
\label{sec:NRQCD}%
The NRQCD factorization formula expresses the annihilation contribution
to the hadronic width of a heavy quarkonium state as an infinite sum of
products of short-distance coefficients, which can be calculated as power
series in $\alpha_s$, and nonperturbative long-distance factors
\cite{Bodwin:1994jh}.  The long-distance factors can be expressed as
expectation values of local four-quark operators 
$\mathcal{O}_c (^{2S+1}L_J)$ that are defined in Ref.~\cite{Bodwin:1994jh}.
These NRQCD matrix elements scale as definite powers of the velocity $v$
of the heavy quark in the quarkonium rest frame.  For each of the
$P$-wave states, there are only two matrix elements that contribute up
to corrections of relative order $v^2$:
$\langle \mathcal{O}_1(^3P_J) \rangle_{\chi_{bJ}}$ and 
$\langle \mathcal{O}_8(^3S_1) \rangle_{\chi_{bJ}}$ for $\chi_{bJ}$ and
$\langle \mathcal{O}_1(^1P_1) \rangle_{h_b}$ and 
$\langle \mathcal{O}_8(^1S_0) \rangle_{h_b}$ for $h_b$.
Heavy-quark spin symmetry can be used to reduce all these matrix elements 
at leading order in $v$ to two independent matrix elements
that we will denote by 
$\langle \mathcal{O}_1 \rangle_{\chi_b}$ and 
$\langle \mathcal{O}_8 \rangle ^{(\Lambda)}_{\chi_b}$:
\begin{subequations}
\begin{eqnarray}
\langle \mathcal{O}_1 \rangle_{\chi_b}  &=&  
\langle \mathcal{O}_1(^1P_1) \rangle_{h_b} \approx
\langle \mathcal{O}_1(^3P_J) \rangle_{\chi_{bJ}},  
\\
\langle \mathcal{O}_8\rangle_{\chi_b}^{(\Lambda)}  &=&  
\langle \mathcal{O}_8(^1S_0)\rangle_{h_b}^{(\Lambda)} \approx
\langle \mathcal{O}_8(^3S_1)\rangle_{\chi_{bJ}}^{(\Lambda)}.  
\end{eqnarray}
\label{O1-O8}%
\end{subequations}
The superscript $(\Lambda)$ on $\langle \mathcal{O}_8 \rangle_{\chi_b}^{(\Lambda)}$
indicates the sensitivity of this matrix element to the NRQCD factorization scale.
There is a total of 10 independent matrix elements that contribute
through order $v^2$~\cite{Huang:1997nt}.

The NRQCD factorization formulas for the annihilation widths of
the $\chi_{bJ}$ at leading order in $v$ can be expressed as
\begin{equation}
\Gamma[\chi_{bJ} \to X] =
A_J(\Lambda) \, \frac{\langle \mathcal{O}_1 \rangle_{\chi_b}}{m_b^4} 
+ A_{8} \, \frac{\langle \mathcal{O}_8 \rangle_{\chi_b}^{(\Lambda)}}{m_b^2} ,
\label{NRQCD:chi}%
\end{equation}
where $X$ represents all possible states that consist of hadrons 
lighter than the $B$ meson, and
$\Lambda$ is the NRQCD factorization scale. 
An analogous equation holds for the rate $d\Gamma[\chi_{bJ} \to X]$ that
is differential in the kinematic variables.
The short-distance coefficients whose leading terms are 
order $\alpha_s^2$ are
\begin{subequations}
\begin{eqnarray}
A_0 &=& \frac{3 C_F}{N_c} \pi \alpha_s^2, 
\label{A0}%
\\
A_2 &=& \frac{4 C_F}{5 N_c} \pi \alpha_s^2, 
\label{A2}%
\\
A_8 &=& \frac{1}{3} n_f \pi \alpha_s^2,
\label{A8}%
\end{eqnarray}
\label{AB:alphas2}%
\end{subequations}
where $N_c=3 $ is the number of colors, $C_F = (N_c^2-1)/(2N_c)=4/3$, 
$n_f = 4$ is the number of light flavors of quarks, including charm,
and the masses of the light quarks have been neglected.
The coefficients  $A_0$ and $A_2$ were first calculated by 
Barbieri, Gatto, and Kogerler in 1976 \cite{Barbieri:1975am}.
The coefficient  $A_8$  was first calculated 
for massless quarks in Ref.~\cite{Bodwin:1994jh}. 
The short-distance coefficients whose leading terms are 
order $\alpha_s^3$ are
\begin{equation}
A_1(\Lambda) =
\frac{C_F  \alpha_s^3}{N_c}
\left[ \left(\frac{587}{54} - \frac{317}{288}   \pi^2\right) C_A
 + \left( - \frac{16}{27} - \frac{4}{9} \log \frac{\Lambda}{2 m_b} \right) 
 		n_f \right] , 
\label{A1}%
\end{equation}
where $C_A=N_c=3$, and, again, the masses of the quarks, including 
the charm quark, have been neglected.
The coefficient $A_1$ was calculated in 
Refs.~\cite{Petrelli:1997ge,Huang:1996cs}. The coefficients 
$A_J(\Lambda)$ depend on $\Lambda$, beginning at order $\alpha_s^3$, 
in such a way as to cancel the dependence of the matrix element 
$\langle \mathcal{O}_8 \rangle^{(\Lambda)}_{\chi_b}$ on $\Lambda$.
The next-to-leading-order terms in the coefficients $A_0$, $A_2$, and $A_8$ 
have been calculated by Petrelli, Cacciari, Greco, Maltoni, and 
Mangano \cite{Petrelli:1997ge} and by Huang and Chao \cite{Huang:1996cs}.
The short-distance coefficients are insensitive to $m_c$, the mass of the 
charm quark. The dependence of the leading terms in $A_J$ and 
$A_8$ on $m_c$ will be calculated in Sec.~\ref{sec:rate}.  
The leading correction term in $A_8$ is proportional 
to $\alpha_s^2 (m_c/m_b)^{4}$.  The leading correction terms in $A_J$ 
are proportional to $\alpha_s^3 (m_c/m_b)^2$.

In the NRQCD factorization formula in Eq.~(\ref{NRQCD:chi}),
the decay rates are summed over all light hadronic states.
In most cases, there are no factorization formulas for less 
inclusive decay rates.  An exception is the inclusive charm decay rate.
The decay of $\chi_{bJ}$ into a final state that includes charm hadrons 
requires the annihilation of the $b \bar b$ pair into partons that 
include a $c \bar c$ pair.  The mass of the charm quark is large 
enough that the contribution to the  short-distance coefficients
from $b \bar b$ annihilation into $c \bar c$ pairs may be calculable in
perturbation theory.  
At leading order in $v$, the NRQCD factorization 
formula for the inclusive charm decay rate of $\chi_{bJ}$  
involves the same matrix elements as the completely inclusive annihilation 
decay rate in Eq.~(\ref{NRQCD:chi}):
\begin{equation}
\Gamma[\chi_{bJ} \to c + X] =
A_J^{(c)}(\Lambda) \, \frac{\langle \mathcal{O}_1 \rangle_{\chi_b}}{m_b^4} 
+ A_8^{(c)} \, \frac{\langle \mathcal{O}_8 \rangle_{\chi_b}^{(\Lambda)}}{m_b^2} ,
\label{Gam-c}%
\end{equation}
where $c + X$ represents all possible states that include a charm hadron.
The short-distance coefficients $A_J^{(c)}$ and $A_8^{(c)}$ 
are power series in $\alpha_s$ whose coefficients are functions 
of the mass ratio $m_c/m_b$. 
We can deduce the limit as $m_c \to 0$ of the 
leading term in $A_8^{(c)}$ from the value of $A_8$ in Eq.~(\ref{A8}): 
$A_8^{(c)} \to (1/3) \pi \alpha_s^2$.  Unlike the coefficients 
in the fully inclusive factorization formula in Eq.~(\ref{NRQCD:chi}), 
the coefficients $A_J^{(c)}$ and $A_8^{(c)}$ in Eq.~(\ref{Gam-c}) 
are sensitive to the charm-quark mass. 
The leading terms in $A_J^{(c)}$ and $A_8^{(c)}$ will be calculated in 
Sec.~\ref{sec:rate}.  We will find that the leading term in $A_J^{(c)}$, 
which is of order $\alpha_s^3$, depends logarithmically on $m_c/m_b$.

The NRQCD matrix elements in Eqs.~(\ref{O1-O8}) can, in principle, 
be calculated by using lattice simulations of NRQCD.  
The feasibility of such calculations was first 
demonstrated by Bodwin, Sinclair, and Kim using quenched lattice 
NRQCD \cite{Bodwin:1996tg}.  
The best calculations available to date have been carried out
using two dynamical light quarks \cite{Bodwin:2001mk}. 
After extrapolation to three light-quark flavors~\cite{Bodwin:2001mk},
the values for the $1P$ multiplet are 
\begin{subequations}
\begin{eqnarray}
\langle \mathcal{O}_1 \rangle_{\chi_b(1P)}  &=& 
3.2 \pm 0.7 \ \textrm{GeV}^{5},
\\
\frac{
\langle \mathcal{O}_8 \rangle_{\chi_b(1P)}^{(\Lambda)}
     }{
\langle \mathcal{O}_1 \rangle_{\chi_b(1P)}
     }  &=& 
0.0021 \pm 0.0007 \ \textrm{GeV}^{-2}.
\label{O8:lattice}%
\end{eqnarray}
\label{O1-O8:lattice}%
\end{subequations}
We have estimated the errors for the three-flavor case by treating
the systematic errors from the quenched and two-flavor calculations as
100\% correlated, treating the statistical errors as uncorrelated, and
adding the resulting systematic and statistical errors for the
three-flavor case in quadrature.
The matrix element $\langle \mathcal{O}_8 \rangle_{\chi_b(1P)}^{(\Lambda)}$
in Eq.~(\ref{O8:lattice}) was computed at $\Lambda=$4.3 GeV.
  
The color-singlet matrix elements can also be estimated by using potential
models for heavy quarkonium:
\begin{equation}
\langle \mathcal{O}_1 \rangle_{\chi_b(nP)} \approx 
\frac{3 N_c}{2 \pi} ~ |R'_{nP}(0)|^2,
\end{equation}
where $N_c = 3$ is the number of colors
and $R_{nP}(r)$ is the radial wave function for the $nP$
multiplet.  The values of $|R'_{nP}(0)|^2$ for four potential models
have been tabulated in Ref.~\cite{Eichten:1995ch}. 
Using the value of $|R'_{nP}(0)|^2$ for the Buchm\"uller-Tye potential, 
we obtain
\begin{subequations}
\begin{eqnarray}
\langle \mathcal{O}_1 \rangle_{\chi_b(1P)} &\approx& 2.03 ~ \textrm{GeV}^5 ,
\label{O11P}%
\\
\langle \mathcal{O}_1 \rangle_{\chi_b(2P)} &\approx& 2.37 ~ \textrm{GeV}^5.
\label{O12P}%
\end{eqnarray}
\label{O1:potmod}%
\end{subequations}
The values of $\langle \mathcal{O}_1 \rangle_{\chi_b(nP)}$ from the four
potential models in Ref.~\cite{Eichten:1995ch} range from those in 
Eqs.~(\ref{O1:potmod}) to those for the Cornell potential, 
which are about 50\% larger.  In the case of $S$-wave states,
there has been recent progress in determining the color-singlet 
NRQCD matrix element from potential models \cite{Bodwin:2006dn}. 
The value of the radial wavefunction at the origin $|R_{1S}(0)|^2$ 
of the $\Upsilon(1S)$ that follows from these methods agrees most closely
with that from the Buchm\"uller-Tye potential.
  
 With the choice of normalization of the 
operators in Ref.~\cite{Bodwin:1994jh}, the color-octet matrix element 
$\langle \mathcal{O}_8 \rangle_{\chi_b}$ can
be interpreted intuitively as the probability density at the origin for
the $b \bar b$ pair to be in a color-octet state. 
One can obtain an order-of-magnitude estimate of
a lower bound on
the  quantity $\langle \mathcal{O}_8 \rangle_{\chi_b}$ 
by using the renormalization properties of the
operators \cite{Bodwin:1994jh}. 
The operator $\mathcal{O}_8$ depends on a renormalization scale $\Lambda$, 
and it mixes under renormalization with $\mathcal{O}_1$. The solution 
to the renormalization group equation at leading order in $\alpha_s$ 
is \cite{Bodwin:1994jh}
\begin{equation}
\langle \mathcal{O}_8\rangle_{\chi_b}^{(m_b)} =
\langle \mathcal{O}_8\rangle_{\chi_b}^{(\Lambda)} 
+ \frac{4 C_F}{3 N_c \beta_0} 
\log \left( \frac{\alpha_s(\Lambda)}{\alpha_s(m_b)} \right)
\frac{\langle \mathcal{O}_1 \rangle_{\chi_b}}{m_b^2} ,
\label{RGsol}%
\end{equation}
where $\beta_0 = (11 N_c - 2 n_f)/6 = 25/6$ 
is the first coefficient in the beta function for QCD with 
$n_f =4$ flavors of light quarks.
The second term on the right side of Eq.~(\ref{RGsol}) 
has a physical interpretation as a
correction from gluon radiation 
that arises from gluon energies between $\Lambda$ and $m_b$.
Since $m_b v$ is the typical momentum scale in a quarkonium state,
we choose $\Lambda=m_b v$. 
Unless there is a near cancellation between the two terms in 
Eq.~(\ref{RGsol}) for $\Lambda = m_b v$,
the matrix element $\langle \mathcal{O}_8\rangle_{\chi_b}^{(m_b)}$
should either be comparable to or larger than
the second term on the right side.  This gives us an order-of-magnitude 
estimate of a lower bound on the matrix element:
\begin{eqnarray}
\langle \mathcal{O}_8\rangle_{\chi_b}^{(m_b)} &\gtrsim&
 \frac{32}{225} \log \left( \frac{\alpha_s(m_b v)}{\alpha_s(m_b)} \right)
\frac{\langle \mathcal{O}_1 \rangle_{\chi_b}}{m_b^2} .
\label{O8lb}%
\end{eqnarray}
Since the one-loop bottom-quark pole mass is 
$m_b^{\textrm{(pole)}}\approx 4.6$ GeV,
we set $m_b = 4.6$ GeV and $m_b v = 1.5$ GeV. Then our 
estimated lower bound on the
dimensionless ratio of the matrix elements 
\begin{equation}
\rho_8 = m_b^2 \langle \mathcal{O}_8  \rangle_{\chi_b}^{(m_b)}/
\langle \mathcal{O}_1 \rangle_{\chi_b}
\label{rho8}%
\end{equation}
is $\rho_8\gtrsim 0.068$.
In comparison, the lattice results in Eq.~(\ref{O1-O8:lattice}),
taken with $m_b=4.6$ GeV, give $\rho_8=0.044\pm 0.015$.
Given the errors, this result is compatible with the estimated 
lower bound from Eq.~(\ref{O8lb}).

In NRQCD, there is no general relation between the matrix elements
$\langle \mathcal{O}_1 \rangle_{\chi_b}$ and
$\langle \mathcal{O}_8  \rangle_{\chi_b}^{(m_b)}$
for different $P$-wave multiplets.
However, if the scale $m_b v^2$ is below the QCD scale 
$\Lambda_{\textrm{QCD}}$, then 
the ratio $\rho_8$ in Eq.~(\ref{rho8})
is the same for all the $P$-wave multiplets~\cite{Brambilla:2001xy}.

\section{Charm quark production in $\bm{\chi_b}$ decay}
\label{sec:momentum}%
\subsection{Perturbative matching}
The coefficients in the NRQCD factorization formula for inclusive charm
production in Eq.~(\ref{Gam-c}) are short-distance quantities that are 
insensitive to the long-distance behavior of the external $b\bar b$ 
states. This implies that the short-distance coefficients can be 
computed in perturbation theory. 
It also implies that, for purposes of computing the
short-distance coefficients, we can replace the external $b\bar b$
hadronic states in the factorization formula with perturbative $b\bar b$
states. We compute the short-distance coefficients by matching the
perturbative expressions for the $b\bar b$ annihilation rates in full
QCD with the corresponding perturbative NRQCD factorization expressions
for the annihilation rates.
The perturbative analog of the NRQCD factorization formula  in 
Eq.~(\ref{Gam-c}) for the annihilation rates of appropriate 
$b \bar b$ states is
\begin{equation}
d\Gamma[b \bar b \to c + X] =
\sum_{J=0}^2 dA_J^{(c)}(\Lambda) \, \frac{\langle \mathcal{O}_1(^3P_J) \rangle_{b \bar b}}{m_b^4} 
+ dA_8^{(c)} \, 
\frac{\langle \mathcal{O}_8(^3S_1) \rangle^{(\Lambda)}_{b \bar b}}{m_b^2} .
\label{Gam-bbbartoc}%
\end{equation}
We have written the factorization formula in differential form so that we can consider 
distributions in kinematic variables associated with the charm quark. 
We can determine the  four short-distance coefficients 
$dA_J^{(c)}$ and $dA_8^{(c)} $  by 
(i) calculating the annihilation rate in perturbative QCD for a $b \bar b$ pair in 
four appropriate independent $b \bar b$ states, 
(ii) calculating the NRQCD matrix elements for each of those four states using 
perturbative NRQCD, 
and then (iii) solving the linear set of equations for the coefficients. 

We wish to calculate the short-distance coefficients at leading order in
$\alpha_s$, which is order $\alpha_s^2$ for $A_8^{(c)}$ and order
$\alpha_s^3$ for $A_J^{(c)}$. At this order, we must take into account
the renormalization of the NRQCD matrix element  
$\langle\mathcal{O}_8(^3S_1)\rangle_{b\bar b}$.  
We regularize the NRQCD matrix element by using  dimensional regularization
in $d = 4-2 \epsilon$ space-time dimensions, and we define the renormalized
NRQCD matrix element by using the modified minimal subtraction 
($\overline{\textrm{MS}}$) prescription. The relation between the bare
operator  $\mathcal{O}_8(^3S_1)$ and the renormalized operator $\mathcal{
O}_8(^3S_1)^{( \Lambda)}$  with NRQCD factorization scale $\Lambda$
is~\cite{Bodwin:1994jh,Petrelli:1997ge}
\begin{equation}
\mathcal{O}_8(^3S_1) =
\mathcal{O}_8(^3S_1)^{(\Lambda)} 
 + \frac{(4 \pi e^{- \gamma})^\epsilon}{\epsilon_{\textrm{UV}}}
 \frac{2 C_F \alpha_s}{3 \pi N_c m_b^2}
 \sum_{J=0}^2 \mathcal{O}_1(^3P_J)   + \ldots .
\label{O8-renorm}%
\end{equation}
The subscript $\textrm{UV}$ indicates that the pole in $\epsilon$ is
associated with an ultraviolet divergence. We have shown  explicitly
only those terms that contribute through order $\alpha_s$ and at
leading order in $v$ to the expectation values in a color-singlet
$P$-wave $b \bar b$ state or in a color-octet $S$-wave $b \bar b$ state.

The perturbative matrix elements of the NRQCD operators regularized with
dimensional regularization are particularly simple if we also use
dimensional regularization to regularize infrared divergences and we
expand the matrix elements in powers of the relative momentum ${\bf q}$
of the $b$ and $\bar b$. In this case, all loop corrections to the
regulated matrix element vanish because there is no scale for the
dimensionally regularized integrals. In particular, the ultraviolet
poles in $\epsilon$ cancel the infrared poles in $\epsilon$. Thus, we
have
\begin{equation}
\langle\mathcal{O}_8(^3S_1)\rangle_{b\bar b}^\textrm{(reg)} =
\langle\mathcal{O}_8(^3S_1)\rangle_{b\bar b}^\textrm{(tree)},
\label{reg-tree}%
\end{equation}
where $\langle \mathcal{O}_8(^3S_1) \rangle^{\textrm{(reg)}}_{b \bar
b} $ is the matrix element of the bare NRQCD operator with 
both infrared and ultraviolet divergences dimensionally regulated, and
$\langle \mathcal{O}_8(^3S_1) \rangle^{\textrm{(tree)}}_{b \bar
b} $ is the tree-level approximation to the matrix element of the bare 
NRQCD operator. If we take the expectation value of Eq.~(\ref{O8-renorm}) in a 
$b\bar b$ state, dimensionally regulating both UV and IR divergences, and 
substitute (\ref{reg-tree}), we find that 
\begin{equation}
\langle \mathcal{O}_8(^3S_1)\rangle_{b \bar b}^{(\Lambda)}  
=
\langle \mathcal{O}_8(^3S_1) \rangle^{\textrm{(tree)}}_{b \bar b} 
- \frac{(4 \pi e^{- \gamma})^\epsilon}{\epsilon_{\textrm{IR}}}
 \frac{2 C_F \alpha_s}{3 \pi N_c m_b^2}
 \sum_{J=0}^2  \langle \mathcal{O}_1(^3P_J)  \rangle_{b \bar b}
 + \ldots.
\label{O8-renorm-bbbar}%
\end{equation}
The subscript $\textrm{IR}$ indicates that the pole in $\epsilon$ is now
associated with an infrared divergence. 

To determine the four short-distance coefficients $A_J^{(c)}$ and $A_8^{(c)} $ in 
Eq.~(\ref{Gam-c}), we must calculate the annihilation rate for four appropriate 
$b \bar b$ states using perturbative QCD.  A convenient choice for these states consists of 
a $b \bar b$ pair in a color-octet $^3S_1$ state, which we denote by $b \bar b_8(^3S_1)$,
and a $b \bar b$ pair in each 
of the three color-singlet $^3P_J$ states, which we denote by $b \bar b_1(^3P_J)$. 
For these states, the factorization formula in Eq.~(\ref{Gam-bbbartoc}) reduces 
at leading order in $\alpha_s$ and at leading order in $v$ to
\begin{subequations}
\begin{eqnarray}
d\Gamma[b \bar b_8(^3S_1)  \to c + X] &=&
 dA_8^{(c)} \, 
\frac{\langle \mathcal{O}_8(^3S_1) \rangle^{(\Lambda)}_{b \bar b_8(^3S_1)}}
     {m_b^2} ,
\label{Gam-bbbar8S}%
\\
d\Gamma[b \bar b_1(^3P_J) \to c + X] &=&
dA_J^{(c)}(\Lambda) \, 
\frac{\langle \mathcal{O}_1(^3P_J) \rangle_{b \bar b_1(^3P_J)}}{m_b^4} 
+ dA_8^{(c)} \, 
\frac{\langle \mathcal{O}_8(^3S_1) \rangle^{(\Lambda)}_{b \bar b_1(^3P_J)}}
{m_b^2} .
\label{Gam-bbbar1P}%
\end{eqnarray}
\end{subequations}
In Eq.~(\ref{Gam-bbbar8S}), the term involving
the color-singlet operator does not contribute 
because it is of higher order in $\alpha_s$.
In Eq.~(\ref{Gam-bbbar1P}), the color-octet matrix element 
$\langle \mathcal{O}_8(^3S_1)\rangle^{(\Lambda)}_{b \bar b_1(^3P_J)}$
can be simplified by using the fact that the tree-level term 
in Eq.~(\ref{O8-renorm-bbbar}) 
does not contribute.  The factorization formula in Eq.~(\ref{Gam-bbbar1P}) 
can then be reduced to
\begin{equation}
d\Gamma[b \bar b_1(^3P_J) \to c + X] =
\left( dA_J^{(c)}(\Lambda) 
- \frac{(4 \pi e^{- \gamma})^\epsilon}{\epsilon_{\textrm{IR}}}
 \frac{2 C_F \alpha_s}{3 \pi N_c} \; dA_8^{(c)} \right)
\frac{\langle \mathcal{O}_1(^3P_J) \rangle_{b \bar b_1(^3P_J)}}{m_b^4}  .
\label{Gam-bbba1P:simp}%
\end{equation}

Eqs.~(\ref{Gam-bbbar8S}) and (\ref{Gam-bbba1P:simp}) can be solved to
obtain the short-distance coefficients $dA_8^{(c)}$ and $dA_J^{(c)}$ in
terms of the perturbative decay rates $d\Gamma[b \bar b_8(^3S_1) \to c +
X]$ and $d\Gamma[b \bar b_1(^3P_J) \to c + X]$ and the perturbative
matrix elements $\langle \mathcal{O}_8(^3S_1) \rangle^{(\Lambda)}_{b
\bar b_8(^3S_1)}$ and $\langle \mathcal{O}_1(^3P_J) \rangle_{b \bar
b_1(^3P_J)}$. At the order in $\alpha_s$ of the present calculation, the
perturbative matrix elements can be computed at tree level. In the next
three subsections, we compute the required perturbative decay rates and
perturbative matrix elements.  As we will see, $d\Gamma[b \bar
b_1(^3P_J) \to c + X]$ contains an infrared divergence that is canceled
by the explicit infrared divergence in the second term on the right side
of Eq.~(\ref{Gam-bbba1P:simp}). The short-distance coefficients are then
infrared finite, as expected.

\subsection{Amplitudes for $\bm{b} \bar{\bm{b}}$ annihilation into charm}

The momenta of the $b$ and $\bar{b}$ that annihilate to produce charm
 can be expressed as
\begin{subequations}
\begin{eqnarray}
p&=&\tfrac{1}{2}P+q,
\\
\bar{p}&=&\tfrac{1}{2}P-q,
\end{eqnarray}
\end{subequations}
where $P$ and $q$ are the total and relative momenta of the $b\bar{b}$ pair.
In the rest frame of the $b\bar{b}$ pair, the explicit 
momenta are $P=(2E_b,0)$ and $q=(0,\bm{q})$, where 
$E_b=\sqrt{m_b^2+\bm{q}^2}$ and $m_b$ is the mass 
of the bottom quark.  An annihilation amplitude can be expressed 
in the form
\begin{equation}
\bar{v}(\bar{p}) \mathcal{A}  u(p) = 
\textrm{Tr} \big[ \mathcal{A} \, u(p) \bar{v}(\bar{p}) \big], 
\label{vbarAu}%
\end{equation}
where $\mathcal{A}$ is a matrix that acts on spinors with both 
Dirac and color indices.
The amplitude in Eq.~(\ref{vbarAu})
can be projected into a particular spin 
and color channel by replacing $u(p)\bar{v}(\bar{p})$ with a projection 
matrix. The color projectors $\pi_1$ and $\pi_8^a$ onto a color-singlet 
state and onto a color-octet state with color index $a$ are
\begin{subequations}
\begin{eqnarray}
\pi_1&=&\frac{1}{\sqrt{N_c}} \mathbbm{1},
\\
\pi_8^a &= &\sqrt{2} \, T^a,
\end{eqnarray}
\label{color-projector}%
\end{subequations}
where $\mathbbm{1}$ is the $3\times 3$ unit matrix 
and $T^a$ is a generator of the fundamental representation of SU(3).
The color projectors are normalized so that
Tr[$\pi_1 \pi_1^\dagger$]=1 and 
Tr[$\pi_8^a \pi_8^{b\dagger}]=\delta^{ab}$.
The projector onto a spin-triplet state with four-momentum $P^\mu$,
rest energy $\sqrt{P^2}=2E_b$,
and spin polarization vector
$\epsilon_S$ satisfying $P \cdot \epsilon_S=0$
is $\epsilon_{S\mu} \Pi_3^\mu$~\cite{%
Kuhn:1979bb,Guberina:1980dc,Bodwin:2002hg}, where
\begin{equation}
\Pi_3^\mu=
\frac{-1}{4\sqrt{2}E_b(E_b+m_b)}
(/\!\!\!{p}+m_b)(\,/\!\!\!\!P\!+\!2E_b)\gamma^\mu
(/\!\!\!\bar{p}-m_b) .
\label{spin-projector}%
\end{equation}
The spin projector is normalized so that
\begin{equation}
\textrm{Tr}[(\epsilon_S \cdot \Pi_3) (\epsilon_S \cdot \Pi_3)^\dagger] 
= 4 p_0 \bar p_0 .
\end{equation}

At leading order in $v$, the amplitude for the annihilation of a $b \bar
b$ pair in a color-octet spin-triplet $S$-wave state with spin
polarization vector $\epsilon_S$ is $\epsilon_{S\mu} \mathcal{A}_8^{a
\mu}$, where
\begin{equation}
\mathcal{A}_8^{a \mu} =
\textrm{Tr} \big[ \mathcal{A} \, (\Pi_3^\mu \otimes \pi_8^a) \big] \Big|_{q=0}.
\label{amp-S}%
\end{equation}
The leading color-octet mechanism for producing charm in $b \bar b$ annihilation
is via the process $b \bar b \to c \bar c$,
whose rate is of order $\alpha_s^2$.
The matrix $\mathcal{A}$ for the process 
$b(p) \bar{b}(\bar{p}) \to  c(p_1) \bar{c}(p_2)$ is 
\begin{equation}
\mathcal{A}[b\bar{b}\to c\bar{c}]=
\frac{-g_s^2}{(p_1+p_2)^2} \bar{u}(p_1) T^b\gamma_\nu v(p_2) \, 
\big[ T^b \gamma^\nu \big].
\end{equation}
Using Eq.~(\ref{amp-S}), we find that the coefficient of  $\epsilon_{S\mu} $
in the annihilation amplitude is
\begin{equation}
\mathcal{A}_8^{a\mu} =
\frac{g_s^2}{2m_b} \bar{u}(p_1) T^a \gamma^\mu v(p_2) ,
\label{amp-8S}%
\end{equation}
where we have omitted terms proportional to $P^\mu$
because $P \cdot \epsilon_S = 0$.

At leading order in the relative velocity $v$ of the $b$ or $\bar b$ in
the quarkonium rest frame, the amplitude for the annihilation of a $b
\bar b$ pair in a color-singlet spin-triplet  $P$-wave state with spin
polarization vector $\epsilon_S$ and orbital-angular-momentum
polarization vector $\epsilon_L$ is $\epsilon_{L\nu} \epsilon_{S\mu}
\mathcal{A}_1^{\mu\nu}$, where 
\begin{equation}
\mathcal{A}_1^{\mu\nu} =
\frac{\partial \ }{\partial q_\nu}
\textrm{Tr} \big[ \mathcal{A} \, (\Pi_3^\mu \otimes \pi_1) \big] \big|_{q=0}.
\label{amp-P}%
\end{equation}
The leading color-singlet mechanism for producing charm in 
$b \bar b$ annihilation is the process $b \bar b \to c \bar c g$,
whose rate is of order $\alpha_s^3$.
The matrix $\mathcal{A}$  for the process 
$b(p) \bar{b}(\bar{p}) \to  c(p_1) \bar{c}(p_2) g(p_3)$ is 
\begin{eqnarray}
\mathcal{A}[b \bar{b} \to  c \bar{c} g] &=&
\frac{-g_s^3}{(p_1+p_2)^2}
  \bar{u}(p_1) T^a \gamma_\lambda v(p_2)\epsilon^{b *}_\sigma(p_3) 
\nonumber
\\
&& \hspace{2cm}
\times \big[ T^aT^b \gamma^\lambda \Lambda(p-p_3) \gamma^\sigma
+ T^bT^a \gamma^\sigma \Lambda(-\bar{p}+p_3)\gamma^\lambda \big],
\label{amp-1P}%
\end{eqnarray}
where $\Lambda(k)$ is defined by
\begin{equation}
\Lambda(k)=\frac{/\!\!\!k+m_b}{k^2-m_b^2}.
\end{equation}
Using Eq.~(\ref{amp-P}),  we find that
\begin{eqnarray}
\mathcal{A}_1^{\mu \nu} &=&
 \frac{-g_s^3}{2 \sqrt{N_c}(P-p_3)^2}
  \bar{u}(p_1) T^a\gamma_\lambda  v(p_2)\epsilon^{a *}_\sigma(p_3)
\nonumber
\\
&& \hspace{3cm} \times
\frac{\partial \ }{\partial q_\nu}
\textrm{Tr} \big\{ \big[\gamma^\lambda \Lambda(p-p_3) \gamma^\sigma
+ \gamma^\sigma \Lambda(-\bar{p}+p_3) \gamma^\lambda\big] \Pi_3^\mu \big\} \big|_{q=0}.
\label{A-1P}%
\end{eqnarray}

\subsection{Color-octet short-distance coefficient}
\label{sec:8match}%
We proceed to calculate the differential coefficient $dA_8^{(c)}$ of the 
color-octet term in the NRQCD factorization formula. 
We use the perturbative factorization formula in Eq.~(\ref{Gam-bbbar8S}),
which requires calculating the annihilation rate of a $b \bar b$ pair 
in a color-octet $^3S_1$ state.  The resulting expression for $dA_8^{(c)}$
will also be needed in the determination of the coefficients $dA_J^{(c)}$
that makes use of the perturbative factorization formula in 
Eq.~(\ref{Gam-bbba1P:simp}).
In that equation, $dA_8^{(c)}$ is multiplied by a pole in $\epsilon$.
It is therefore necessary to calculate  $dA_8^{(c)}$ in $d= 4 - 2 \epsilon$ 
space-time dimensions.

The differential annihilation rate of a color-octet $^3S_1$ $b \bar b$ state 
into charm through the color-octet process $ b \bar b \to c \bar c$
can be expressed in the form
\begin{equation}
d \Gamma [b \bar b_8(^3S_1) \to c + X] =
\left( \frac{1} {d-1} I_{\mu\alpha} 
 \sum_{c \bar c} 
 \mathcal{A}_8^{a \mu} \mathcal{A}_8^{a \alpha *}
\right)
     \,
 d \Phi_2,
\label{dGamma8}%
\end{equation}
where $\mathcal{A}_8^{a \mu}$ is the amplitude in Eq.~(\ref{amp-8S}),
$d \Phi_2$ is the differential 2-body phase space for $c \bar c$, 
and $I^{\mu\nu}$ is the projection tensor for spin 1:
\begin{equation}
I^{\mu\nu}=-g^{\mu\nu}+\frac{P^\mu P^\nu}{P^2}.
\label{Imunu}%
\end{equation}
The factor of $1/(d-1)$ in Eq.~(\ref{dGamma8})  comes from averaging 
over the spin states of the $b \bar b$ pair.  The explicit sum in 
Eq.~(\ref{dGamma8}) is over the color and spin states of the $c$ and $\bar c$.
The evaluation of that sum gives
\begin{equation}
I_{\mu \alpha} \sum_{c \bar c}
\mathcal{A}_8^{a \mu} \mathcal{A}_8^{a \alpha *}
= (4\pi \alpha_s\Lambda^{2\epsilon})^2 (N_c^2-1)\left( d - 2 + r \right),
\label{sumAA*}%
\end{equation}
where $g_s^2=4\pi\alpha_s\Lambda^{2\epsilon}$ and
$\Lambda$ is the scale associated with dimensional regularization.
In our calculation, $\Lambda$ becomes the NRQCD factorization scale.
We have set $E_b \to m_b$ in $I^{\mu \nu}$ for consistency with the 
prescription for  $\mathcal{A}_8^{a \mu}$ in Eq.~(\ref{amp-S}),
which involved expanding to leading order in $v$. 

We wish to obtain an expression for the coefficient that is differential 
in the energy of the charm quark.  We therefore integrate 
over the entire 2-body phase space, except for $E_1$, the energy of
the charm quark in the $b \bar b$ rest frame.  
In the center-of-momentum frame, the differential 2-body 
phase space in $d=4-2\epsilon$ space-time dimensions reduces to
\begin{equation}
d\Phi_2 =
c_2(\epsilon)
\frac{|\bm{p}_1|^{1- 2\epsilon}}{8 \pi E_b}
\delta(E_1 - E_b) dE_1,
\label{dPS2}%
\end{equation}
where $|\bm{p}_1|=(E_1^2 - m_c^2)^{1/2}$ is the magnitude of the
three-momentum of 
the charm quark and $2 E_b$ is the energy of the $b \bar b$ pair.
The dimensionless coefficient $c_2(\epsilon)$, which reduces to 1 
as $\epsilon\to 0$, is defined by
\begin{equation}
c_2(\epsilon)=
(4\pi)^\epsilon\frac{\Gamma(\tfrac{3}{2})}{\Gamma(\tfrac{3}{2}-\epsilon)}.
\label{c2e}%
\end{equation}
It is useful to express the differential phase space in terms of 
an energy fraction $x_1$ for the charm quark defined by 
\begin{equation}
x_1=E_1/E_b.
\label{x1}%
\end{equation}
There is some ambiguity in the choice of $E_b$.
The choice $E_b = M_{\chi_{bJ}}/2$ gives the correct kinematic
limits on the energy of the charm quark.
However, we choose $E_b=m_b$ in order to maintain consistency with
the nonrelativistic approximation that we used in computing
$\mathcal{A}_8^{a \mu}$ in Eq.~(\ref{amp-S}).
The expression for the differential phase space then  reduces to 
\begin{equation}
d\Phi_2 =
\frac{c_2(\epsilon)}{[(1-r) m_b^2]^\epsilon} \,
\frac{\sqrt{1-r}}{8 \pi}
\delta(1 - x_1) dx_1,
\label{dPS2-m}%
\end{equation}
where $r$ is the square of the ratio of the charm- 
and bottom-quark masses:
\begin{eqnarray}
r= m_c^2/m_b^2.
\label{r:c/b}%
\end{eqnarray}

Inserting the differential phase space in Eq.~(\ref{dPS2-m})
into Eq.~(\ref{dGamma8}) and using Eq.~(\ref{sumAA*}),
we find that
the expression for the differential annihilation rate reduces to
\begin{equation}
d\Gamma[b \bar b_8(^3S_1) \to c+X]
=
\frac{c_2(\epsilon)\Lambda^{4\epsilon}}{[(1-r) m_b^2]^{\epsilon}}
\times
\frac{2 (N_c^2-1)(d-2+r)\sqrt{1-r}\,\pi\alpha_s^2 }{d-1} \,
\delta(1-x_1) dx_1.
\label{Gamma-bbar8}%
\end{equation}
To complete the matching calculation of the coefficient $d A_8^{(c)}$,
we need to evaluate the NRQCD matrix element on the right side of 
the factorization formula in Eq.~(\ref{Gam-bbbar8S}).
The $b \bar b$ states have the standard relativistic normalizations.
At leading order in the nonrelativistic expansion, the matrix element 
is therefore
\begin{equation}
\langle \mathcal{O}_8(^3S_1) \rangle_{b \bar b_8(^3S_1)}   
= 
4 (N_c^2-1) m_b^2.
\label{O83S1-bbbar}
\end{equation}
Inserting Eqs.~(\ref{Gamma-bbar8})  and (\ref{O83S1-bbbar})
into Eq.~(\ref{Gam-bbbar8S}), we find that the differential coefficient 
$d A_8^{(c)}$ in $d$ dimensions is 
\begin{equation}
dA_8^{(c)}
=
\frac{c_2(\epsilon)\Lambda^{4\epsilon}}{[(1-r) m_b^2]^{\epsilon}}
\times
\frac{(d-2+r)\sqrt{1-r}}{2(d-1)} \, \pi\alpha_s^2 \,
\delta(1-x_1) dx_1.
\label{dA8c}%
\end{equation}
Upon setting $\epsilon = 0$, we find that
the differential coefficient with respect to
the energy fraction of the charm quark reduces to
\begin{equation}
\frac{d A_8^{(c)}}{dx_1} =
 \frac{(1 + r/2)\sqrt{1-r}}{3} \pi \alpha_s^2 \, \delta(1- x_1).
\label{dA8cdx1}%
\end{equation}

\subsection{Color-singlet short-distance coefficients}
\label{color-singlet-coeffs}%
We next calculate the differential coefficients $dA_J^{(c)}$ of the 
color-singlet terms in the NRQCD factorization formula. 
We use the perturbative factorization formula in Eq.~(\ref{Gam-bbba1P:simp}),
which requires calculating the annihilation rate of a $b \bar b$ pair 
in a color-singlet $^3P_J$ state for $J=0$, 1, and 2.
This annihilation rate is infrared divergent
at leading order in $\alpha_s$. 
We use dimensional regularization in  $d=4-2\epsilon$ 
space-time dimensions to regularize the infrared divergence.

The differential annihilation rate of a color-singlet $^3P_J$ $b \bar b$ state 
into charm through the color-singlet process $b \bar b \to c \bar c g$
can be expressed in the form
\begin{equation}
d\Gamma[b \bar b_1(^3P_J) \to c+X]=
\left( \frac{1} {S_J(d)} K^J_{\mu\nu;\alpha\beta}
\sum_{c\bar{c}g} \mathcal{A}_1^{\mu\nu} \mathcal{A}_1^{*\alpha\beta} 
 \right)
d\Phi_3 ,
\label{Gamma1}%
\end{equation}
where $\mathcal{A}_1^{\mu \nu}$ is the amplitude in Eq.~(\ref{A-1P}),
$d \Phi_3$ is the differential three-body phase space for $c \bar c g$, 
and the $K^J_{\mu\nu;\alpha\beta}$ are the projection tensors for 
total angular momentum $J$.
In $d$ space-time dimensions, these projectors are~\cite{Petrelli:1997ge}
\begin{subequations}
\begin{eqnarray}
K^0_{\mu\nu;\alpha\beta}&=&
\frac{1}{d-1}I^{\mu\nu}I^{\alpha\beta},
\\
K^1_{\mu\nu;\alpha\beta}&=&
\frac{1}{2}
\left( I^{\mu\alpha}I^{\nu\beta}
      -I^{\mu\beta }I^{\nu\alpha}\right),
\\
K^2_{\mu\nu;\alpha\beta}&=&
\frac{1}{2}
\left( I^{\mu\alpha}I^{\nu\beta}
      +I^{\mu\beta }I^{\nu\alpha}\right)
-\frac{1}{d-1}I^{\mu\nu}I^{\alpha\beta},
\end{eqnarray}
\end{subequations}
where $I^{\mu \nu}$ is given in Eq.~(\ref{Imunu}).
For $J=0$, 1, and 2, $K^{J}_{\mu\nu;\alpha\beta}$ projects the tensor 
$\mathcal{A}_1^{\mu \nu}$ onto its trace, its antisymmetric 
part, and its traceless symmetric part, respectively.
The factor of $S_J(d)$ in Eq.~(\ref{Gamma1}) comes from 
averaging over the angular momentum states of the $b \bar b$ pair.
The spin-$J$ multiplicities in $d$ dimensions are 
\begin{subequations}
\begin{eqnarray}
S_0(d)&=&K^0_{\mu\nu;\alpha\beta} K^{0\,\mu\nu;\alpha\beta}=1,
\\
S_1(d)&=&K^1_{\mu\nu;\alpha\beta} K^{1\,\mu\nu;\alpha\beta}
=\tfrac{1}{2}(d-1)(d-2),
\\
S_2(d)&=&K^2_{\mu\nu;\alpha\beta} K^{2\,\mu\nu;\alpha\beta}
=\tfrac{1}{2}(d+1)(d-2).
\end{eqnarray}
\label{SJ}%
\end{subequations}
The explicit sum in Eq.~(\ref{Gamma1}) is over the color and spin states 
of the $c$, $\bar c$, and $g$.
In the expression for the amplitude $\mathcal{A}_1^{\mu \nu}$ in 
Eq.~(\ref{amp-1P}), the only factors that depend on the 
spins and colors of the $c \bar c g$ are $\bar{u}(p_1)$, 
$v(p_2)$, and $\epsilon^{a *}_\sigma(p_3)$.  
The sum over the spins and colors of the $c \bar c g$ are
\begin{eqnarray}
&& \sum_{c \bar c g} \,
\bar{u}(p_1) T^a \gamma_\lambda v(p_2)\epsilon^{a *}_\sigma(p_3)\,
\big[\bar{u}(p_1) T^b \gamma_\rho v(p_2)\epsilon^{b *}_\tau(p_3) \big]^*
\nonumber
\\
&& \hspace{5cm} = 
-\frac{N_c^2-1}{2} g_{\sigma \tau} \textrm{Tr} 
\big[ (/\!\!\!p_1 + m_c) \gamma_\lambda (/\!\!\!p_2 - m_c) \gamma_\rho \big].
\label{sum:ccbarg}%
\end{eqnarray}
We have omitted terms from the sum over gluon spins that are 
proportional to $p_{3\sigma}$ or $p_{3 \tau}$, because they give zero 
when they are contracted with the trace in Eq.~(\ref{A-1P}) or
its complex conjugate.
After evaluating the Dirac traces in Eq.~(\ref{sum:ccbarg}) and in 
Eq.~(\ref{A-1P}), we reduce the contracted tensors in the differential 
decay rate in Eq.~(\ref{Gamma1}) to complicated functions of
Lorentz scalars, which we will report later in this section.

We wish to obtain expressions for the coefficients $A_J^{(c)}$ that are differential 
in the momentum of the charm quark.  We must therefore integrate 
over the entire three-body phase space, except for the energy $E_1$ of
the charm quark in the $b \bar b$ rest frame.  
The differential three-body phase space in the center-of-momentum frame
in $d=4-2\epsilon$ space-time dimensions is 
computed in Appendix \ref{sec:dphi3}:
\begin{equation}
d\Phi_3 =
\frac{(4\pi)^{2\epsilon}}{\Gamma(2-2\epsilon)}
\delta(E_1 + E_2 + E_3 - 2 E_b)
\frac{dE_1 dE_2 dE_3}
     {32 \pi^3\left[ -\lambda(\bm{p}_1^2,\bm{p}_2^2,\bm{p}_3^2)
 \right]^{\epsilon}} ,
\label{dPS3}%
\end{equation}
where $\lambda(x,y,z) = x^2+y^2+z^2 -2(xy+yz+zx)$
and $|\bm{p}_i| = (E_i^2 - m_i^2)^{1/2}$ is the 
magnitude of the three-momentum of particle $i$.
The physical region of $E_1$, $E_2$, and $E_3$ is determined 
by the delta function 
and by the condition 
$-\lambda(\bm{p}_1^2,\bm{p}_2^2,\bm{p}_3^2) \ge 0$. 
The physical region can be determined from the expression  
\begin{equation}
-\lambda(\bm{p}_1^2,\bm{p}_2^2,\bm{p}_3^2)
=
(|\bm{p}_1|+|\bm{p}_2|+|\bm{p}_3|)
(|\bm{p}_1|+|\bm{p}_2|-|\bm{p}_3|)
(|\bm{p}_2|+|\bm{p}_3|-|\bm{p}_1|)
(|\bm{p}_3|+|\bm{p}_1|-|\bm{p}_2|).
\end{equation}
We let the energies of the $c$, $\bar c$, and $g$ be  $E_1$, $E_2$, and $E_3$, 
respectively.  It is convenient to introduce dimensionless energy variables 
$x_i$ defined by
\begin{equation}
x_i = E_i/E_b.
\label{xi-def}%
\end{equation}
We can use the delta function in Eq.~(\ref{dPS3}) to integrate over 
$x_2$.  If we set $E_b=m_b$, then the differential phase space for 
$c \bar c g$ reduces to
\begin{equation}
d\Phi_3
=\frac{c_3(\epsilon)}{[(x_1^2 - r)x_3^2 
                    (1-\cos^2\theta_{13}) m_b^4]^{\epsilon}}
\frac{m_b^2}{32\pi^3} dx_1 dx_3,
\label{PS3}%
\end{equation}
where $c_3(\epsilon)$ is defined by 
\begin{equation}
c_3(\epsilon) =
\frac{(4\pi)^{2\epsilon}\, \Gamma(\tfrac{3}{2})}
     {\Gamma(1-\epsilon) \Gamma(\tfrac{3}{2}-\epsilon)}
     =
\frac{(2\pi)^{2\epsilon}}{\Gamma(2-2\epsilon)},
\label{c3e}%
\end{equation}
and $\theta_{13}$ is the angle between the momenta $p_1$ and $p_3$:
\begin{equation}
\sin^2 \theta_{13} =
-\lambda(\bm{p}_1^2,\bm{p}_2^2,\bm{p}_3^2)/(4 \bm{p}_1^2 \bm{p}_3^2).
\end{equation}
The ranges of the variables $x_1$ and $x_3$ are given by
\begin{subequations}
\begin{eqnarray}
\label{x3-range}%
\sqrt{r}\leq &x_1& \leq1
,
\\
\label{x1-range}%
x_3^- \leq &x_3& \leq x_3^+ 
,
\end{eqnarray}
\label{range}%
\end{subequations}
where the endpoints of the $x_3$ integral are 
\begin{equation}
x_3^\pm 
=\frac{2(1-x_1)}{2-x_1 \mp \sqrt{x_1^2-r}}.
\label{x3pm}%
\end{equation}

After integrating over the energy  fractions of the $\bar c$ and $g$,
we find that
the differential annihilation rate in Eq.~(\ref{Gamma1}) reduces to 
\begin{equation}
d\Gamma[b \bar b_1(^3P_J) \to c+X] =
\frac{8 C_F \alpha_s^3\Lambda^{6\epsilon}}{m_b^2}
\left[ c_3(\epsilon) m_b^{-4 \epsilon} \, 
\hat{\Gamma}_{\textrm{div}}^J(x_1)
+
\hat{\Gamma}_{\textrm{fin}}^J(x_1)
\right]dx_1,
\label{Gamma-bbbar1}%
\end{equation}
where the coefficient $c_3(\epsilon)$ is defined in Eq.~(\ref{c3e}).
The dimensionless functions $\hat{\Gamma}_{\textrm{div}}^J(x_1)$ 
are defined by
\begin{subequations}
\begin{eqnarray}
\hat{\Gamma}_{\textrm{div}}^0(x_1)&=&
\frac{(d-2+r)I_0(x_1) - 4 [I_1(x_1) - I_2(x_1)]}
     {(d-1)(x_1^2-r)^{\epsilon}},
\\
\hat{\Gamma}_{\textrm{div}}^1(x_1)&=&
\frac{(d-3)(d-2+r)I_0(x_1) + 4 [I_1(x_1) - I_2(x_1)]}
     {(d-1)(d-2)(x_1^2-r)^{\epsilon}},
\\
\hat{\Gamma}_{\textrm{div}}^2(x_1)&=&
\frac{(d^2-2d-1)(d-2+r)I_0(x_1) - 4(d-3) [I_1(x_1) - I_2(x_1)]}
     {(d-1)(d+1)(d-2)(x_1^2-r)^{\epsilon}},
\end{eqnarray}
\label{div}%
\end{subequations}
where the functions $I_n(x_1)$ are integrals over $x_3$:
\begin{equation}
I_n(x_1)=\int_{x_3^-}^{x_3^+}  dx_3 \frac{(1-x_1)^n}
          {x_3^{n+2+2\epsilon}(1-\cos^2\theta_{13})^\epsilon}.
\label{In}%
\end{equation}
These integrals, which are logarithmically infrared divergent,
are evaluated analytically in Appendix~\ref{sec:In}.
They can be expressed in terms of two distributions that are singular
at $x_1 = 1$:  the Dirac delta function $\delta(1-x_1)$
and a distribution $[1/(1-x_1)]_{\sqrt{r}}$ that is defined by
\begin{equation}
\int_{x}^1g(x_1) [f(x_1)]_{\sqrt{r}} dx_1\equiv
\int_{x}^1[g(x_1)-g(1)] f(x_1) dx_1
- g(1)\int_{\sqrt{r}}^x f(x_1) dx_1
\label{root-def}%
\end{equation}
for any $x$ in the interval $\sqrt{r} \le x < 1$ 
and any smooth function $g(x_1)$.
The dimensionless functions $\hat{\Gamma}_{\textrm{fin}}^J(x_1)$
in Eq.~(\ref{Gamma-bbbar1}) are defined by
\begin{subequations}
\begin{eqnarray}
\hat{\Gamma}_{\textrm{fin}}^0(x_1)&=&
\int_{x_3^-}^{x_3^+}
\frac{ 2(1-x_3)(8+x_3)C(x_1,x_3) + 3r(4-x_3)}
     {12(1-x_3)^2}\, \frac{dx_3}{x_3},
\\
\hat{\Gamma}_{\textrm{fin}}^1(x_1)&=&
-\frac{1}{3}
\int_{x_3^-}^{x_3^+} 
C(x_1,x_3)
\frac{dx_3}{x_3},
\\
\hat{\Gamma}_{\textrm{fin}}^2(x_1)&=& \int_{x_3^-}^{x_3^+}
\frac{(1-x_3)(5+x_3)C(x_1,x_3) + 3r(2-x_3) }
     {15(1-x_3)^2}\, \frac{dx_3}{x_3},
\end{eqnarray}
\label{Gamhat-fin}%
\end{subequations}
where $C(x_1,x_3)$ is the function 
\begin{equation}
C(x_1,x_3)=\frac{(1-x_1)^2+(x_1+x_3-1)^2}{x_3^2}.
\end{equation}
The results from carrying out the integrations over $x_3$ in 
Eqs.~(\ref{div}) and (\ref{Gamhat-fin}) are tabulated in 
Appendix~\ref{app:intx3}.

To complete the matching calculation of the coefficient $d A_J^{(c)}$,
we need to evaluate the NRQCD matrix element on the right side of 
the factorization formula in Eq.~(\ref{Gam-bbba1P:simp}).
The $b \bar b$ states have a nonstandard normalization that corresponds
to the procedure that we used in computing the full QCD rate 
$d\Gamma[b \bar b_1(^3P_J) \to c+X]$  [Eqs.~(\ref{amp-S}) and 
(\ref{Gamma1})].
Application of that procedure in NRQCD
is equivalent to the use of $b \bar b$ states that are normalized to
$3(2E_b)^2/\bm{q}^2$, instead of the conventional $(2E_b)^2$, 
where $\bm{q}$ is the momentum of the $b$ quark in
the quarkonium rest frame. The matrix element at leading order in the
nonrelativistic approximation is then
\begin{equation}
\langle \mathcal{O}_1(^3P_J) \rangle_{b \bar b_1(^3P_J)} 
= 8 N_c m_b^2 .
\label{O1bbbar}%
\end{equation}
Substituting Eqs.~(\ref{dA8c}), (\ref{Gamma-bbbar1}),  
and (\ref{O1bbbar}) into the factorization formula
(\ref{Gam-bbba1P:simp}),
we obtain
\begin{eqnarray}
dA_J^{(c)}(\Lambda) &=&
\frac{C_F\alpha_s^3}{N_c}\,
c_3(\epsilon)\left(\frac{\Lambda^6}{m_b^4}\right)^\epsilon
\bigg\{
\hat{\Gamma}^J_{\textrm{div}}(x_1)
+
\hat{\Gamma}^J_{\textrm{fin}}(x_1)
\nonumber\\
&
+
&
\left[
\frac{1}{\epsilon_{\textrm{IR}}}
+\frac{2(r-1)}{3(2+r)}
+\log \frac{m_b^2}{(1-r)\Lambda^2}
\right]
\frac{(2+r)\sqrt{1-r}}{9}\delta(1-x_1)
\bigg\}\,dx_1
+O(\epsilon),~~~
\label{AJc-formula}%
\end{eqnarray}
where we use
\begin{equation}
(4\pi e^{-\gamma})^\epsilon\frac{c_2(\epsilon)}{c_3(\epsilon)}
=1+O(\epsilon^2).
\end{equation}
The explicit infrared divergence in Eq.~(\ref{AJc-formula}) is 
canceled by the infrared divergence in 
$\hat{\Gamma}_{\textrm{div}}^J(x_1)$.
Therefore the expression in Eq.~(\ref{AJc-formula}) is finite at 
$\epsilon=0$, and so we can neglect the $\epsilon$ dependence in the 
prefactor. The only dependence on the scale $\Lambda$ that remains
appears in the bracket in Eq.~(\ref{AJc-formula}). 

It is now straightforward to determine the coefficients $d A_J^{(c)}$.
Our final results for the differential coefficients with respect to
the energy fraction $x_1$ of the charm quark are 
\begin{subequations}
\begin{eqnarray}
\frac{d A_0^{(c)}(\Lambda)}{dx_1 \ \ } &=&
 \frac{C_F\alpha_s^3}{N_c}
\bigg\{
\left[ \left(
\frac{2(2+r)}{9} \log \frac{4 (1-\sqrt{r}) m_b}{\sqrt{r}\Lambda} 
+\frac{1+r}{9}
\right)\sqrt{1-r} \right.
\nonumber \\
&& \hspace{4cm} \left.
- \frac{4 + 3 r}{18} \log \frac{1+\sqrt{1-r}}{1-\sqrt{1-r}}
\right]
\delta(1-x_1)
\nonumber\\
&&
+
\frac{2}{9}
\left( 28-35x_1
+ (2+r) \Big[ \frac{1}{1-x_1} \Big]_{\sqrt{r}} \right)
\sqrt{x_1^2-r}
\nonumber\\
&&
+\left[
\frac{3}{2}+r-3x_1(1-x_1)
\right]
\log \frac{x_1+\sqrt{x_1^2-r}}{x_1-\sqrt{x_1^2-r}}
\nonumber\\
&&
-\frac{1}{6}
\log\frac{2-x_1+\sqrt{x_1^2-r}}{2-x_1-\sqrt{x_1^2-r}} 
\bigg\}
,
\\
\frac{d A_1^{(c)}(\Lambda)}{dx_1 \ \ } &=&
\frac{C_F\alpha_s^3}{N_c}
\bigg\{
\left[ \left( 
\frac{2(2+r)}{9} \log \frac{4 (1-\sqrt{r}) m_b}{\sqrt{r}\Lambda} 
       +\frac{1}{18}
\right) \sqrt{1-r} \right.
\nonumber \\
&& \hspace{4cm} \left.
- \frac{8 + 3 r}{36} \log \frac{1+\sqrt{1-r}}{1-\sqrt{1-r}}
\right]
\delta(1-x_1)
\nonumber\\
&&
+
\frac{2}{9}
\left( -\frac{1-5x_1}{2} 
+ (2+r) \Big[ \frac{1}{1-x_1} \Big]_{\sqrt{r}} \right)
\sqrt{x_1^2-r}
\nonumber \\
&& -\frac{1}{3}
\log\frac{2-x_1+\sqrt{x_1^2-r}}{2-x_1-\sqrt{x_1^2-r}} 
\bigg\},
\\
\frac{d A_2^{(c)}(\Lambda)}{dx_1 \ \ } &=&
 \frac{C_F\alpha_s^3}{N_c}
\bigg\{
\left[ \left( 
\frac{2(2+r)}{9} \log \frac{4 (1-\sqrt{r}) m_b}{\sqrt{r}\Lambda} 
      +\frac{7 +4 r}{90}
\right)\sqrt{1-r} \right.
\nonumber \\
&& \hspace{4cm} \left.
- \frac{40 + 21 r}{180} \log \frac{1+\sqrt{1-r}}{1-\sqrt{1-r}}
\right]
\delta(1-x_1)
\nonumber\\
&&
+
\frac{2}{9}
\left( \frac{73-89x_1}{10}
+ (2+r) \Big[ \frac{1}{1-x_1} \Big]_{\sqrt{r}} \right)
\sqrt{x_1^2-r}
\nonumber\\
&&
+\frac{2}{5}\big[1+r-2x_1(1-x_1)\big]
\log \frac{x_1+\sqrt{x_1^2-r}}{x_1-\sqrt{x_1^2-r}}
\nonumber\\
&&
-\frac{1}{15}
\log\frac{2-x_1+\sqrt{x_1^2-r}}{2-x_1-\sqrt{x_1^2-r}} 
\bigg\},
\end{eqnarray}
\label{dAJcdx1}%
\end{subequations}
where  $[1/(1-x_1)]_{\sqrt{r}}$ is the distribution defined in
Eq.~(\ref{root-def}).

In Eq.~(\ref{dAJcdx1}), the terms involving the $[1/(1-x_1)]_{\sqrt{r}}$
distribution diverge as $1/(1-x_1)$ as $x_1\to 1$. 
These singularities arises
because, as $x_1\to 1$, the energy of the real gluon in the final state
goes to zero, giving rise to an infrared divergence in the rate. The
second term in the definition of the  $[1/(1-x_1)]_{\sqrt{r}}$
distribution provides a negative contribution that cancels this
divergence when one integrates over a region in $x_1$ that contains
the point $x_1=1$.
Suppose that one integrates over the region $x\leq x_1\leq
1$. Then, owing to the second term in the definition of the 
$[1/(1-x_1)]_{\sqrt{r}}$ distribution (\ref{root-def}), the result is
dominated in the limit $x\to 1$ by a term that is proportional to
$-\log[1/(1-x)]$. Such unphysical divergences are a symptom of the fact
that the perturbation expansion in $\alpha_s$ breaks down in the limit
$x_1\to 1$ because of the appearance of large logarithms of $1-x_1$. 
A correct treatment of the region near $x_1=1$
would involve the resummation of logarithms of $1-x_1$
(Ref.~\cite{Bauer:2000ew,Bauer:2001rh,Fleming:2002rv,
Fleming:2002sr,Fleming:2004rk,Fleming:2004hc}).
As $x\to 1$, real gluon emission is suppressed. Hence, the resummation
of logarithms of $1-x$ generally leads to a Sudakov factor that
suppresses the rate near $x=1$ (Ref.~\cite{Bauer:2000ew,Bauer:2001rh,
Fleming:2002rv,Fleming:2002sr,Fleming:2004rk,Fleming:2004hc}).
Consequently, as $x\to 1$, we expect the resummed distribution to 
turn over, rather than to diverge, and to approach zero smoothly 
at $x=1$. We note that, in the rate integrated over all $x_1$, 
logarithms of $1-x_1$ do not appear, and resummation is not necessary
in order to obtain a reliable result.

In the limit $x\to 1$, the velocity expansion of NRQCD also breaks down
because of kinematic constraints near the energy endpoint
\cite{Beneke:1997qw}. A correct treatment of this problem would
involve the inclusion of shape functions \cite{Beneke:1997qw} for the
$\chi_{bJ}$ mesons. 
In general, the inclusion of shape functions has the effect of smearing
the energy distribution near the end point.
We expect that these smearing effects will
be important for $1-x_1$ less than $v\approx 0.3$. In this region, the
expression in Eq.~(\ref{dAJcdx1}) should not be taken as an accurate
estimate of the distribution. (For a discussion of these effects in the
decay of the $\Upsilon$ meson into a photon plus light hadrons, see
Ref.~\cite{GarciaiTormo:2005ch}.)
In the total rate integrated over $x_1$, the velocity expansion is
well behaved and the effects from the shape function are of higher order
in $v^2$.

The resummation of logarithms of $1-x_1$ and the inclusion of shape
functions are beyond the scope of this paper. In the absence of such
analyses, one should treat our results with caution in the region near
$x_1=1$. 

\subsection{Charm-quark momentum distribution}

The NRQCD factorization formula in Eq.~(\ref{Gam-c}) can be expressed 
in a form that is differential in the energy fraction $x_1$ of the
charm quark:
\begin{eqnarray}
\frac{d\Gamma}{dx_1}[\chi_{bJ} \to c + X] &=&
\frac{dA_J^{(c)}(\Lambda)}{dx_1} \, 
\frac{\langle \mathcal{O}_1 \rangle_{\chi_b}}{m_b^4} 
+ \frac{dA_8^{(c)}}{dx_1} \, 
\frac{\langle \mathcal{O}_8 \rangle_{\chi_b}^{(\Lambda)}}{m_b^2},
\label{dGamdx1}%
\end{eqnarray}
where the color-singlet coefficients $dA_J^{(c)}(\Lambda)/dx_1$
are given in Eqs.~(\ref{dAJcdx1}) and the color-octet coefficient
$dA_8^{(c)}/dx_1$ is given in Eq.~(\ref{dA8cdx1}).

The momentum  distribution for the charm quark can be obtained 
from Eq.~(\ref{dGamdx1}) by a simple change of variables.
It is convenient to express that momentum  in terms of the
fraction $y_1$ of the maximum momentum for a charm quark 
that is kinematically allowed in the annihilation of a $b \bar b$ 
pair at threshold:
\begin{eqnarray}
y_1 = \sqrt{\frac{x_1^2 - r}{1-r}} .
\label{y1-def}%
\end{eqnarray}
The range of $y_1$ is $0 < y_1 < 1$.  
The inverse relation is
\begin{eqnarray}
x_1 = \sqrt{(1-r)y_1^2+r} .
\label{x1-y1}%
\end{eqnarray}
The distribution in the fractional momentum $y_1$ can then 
be written as
\begin{eqnarray}
\frac{d\Gamma}{dy_1} = 
\frac{(1-r) y_1}{\sqrt{(1-r)y_1^2 + r}} \, \frac{d\Gamma}{dx_1}.
\label{dGy1}%
\end{eqnarray}
The singular distribution $[1/(1-x_1)]_{\sqrt{r}}$ in the coefficients 
$d A_J^{(c)}/dx_1$ in Eqs.~(\ref{dAJcdx1})
can be transformed into a singular distribution in the variable $y_1$ as 
follows.
From Eq.~(\ref{root-def}) we can derive the identity
\begin{eqnarray}
\left[
\frac{1}{1-x_1}
\right]_{\sqrt{r}}
dx_1
=
\left\{
h(y_1)
\left[\frac{1}{1-y_1}\right]_+
+\delta(1-y_1)
\int_0^1 dy'\,\frac{h(1)-h(y')}{1-y'}
\right\}dy_1,
\label{identity2}%
\end{eqnarray}
where 
\begin{eqnarray}
h(y_1)=\left(\frac{1-y_1}{1-x_1}\right)\frac{dx_1}{dy_1}.
\end{eqnarray}
Note that
$h(1)=1$. Using Eq.~(\ref{x1-y1}) to compute
$h(y_1)$ and substituting the results into
Eq.~(\ref{identity2}), we obtain 
\begin{eqnarray}
\left[
\frac{1}{1-x_1}
\right]_{\sqrt{r}}
dx_1
=
\bigg\{ \frac{y_1\big[1 + \sqrt{(1-r)y_1^2 + r}\,\big] }
            {(1+y_1)\sqrt{(1-r)y_1^2 + r }}
	 \,\left[\frac{1}{1-y_1}\right]_+ 
+ \log(1 + \sqrt{r})  \delta(1-y_1) 
\bigg\} \,
dy_1,
\nonumber\\
\label{sing-xy}%
\end{eqnarray}
where the plus distribution  $[1/(1-y_1)]_+$ is defined by
\begin{equation}
\int_{y}^1g(y_1) [f(y_1)]_+ dy_1\equiv
\int_{y}^1[g(y_1)-g(1)] f(y_1) dy_1
- g(1)\int_0^y f(y_1) dy_1
\label{plus-def}%
\end{equation}
for any $y$ in the interval $0 \le y < 1$
and any smooth function $g(y_1)$.

\begin{figure}[htb]
\includegraphics*[width=12cm,angle=0,clip=true]{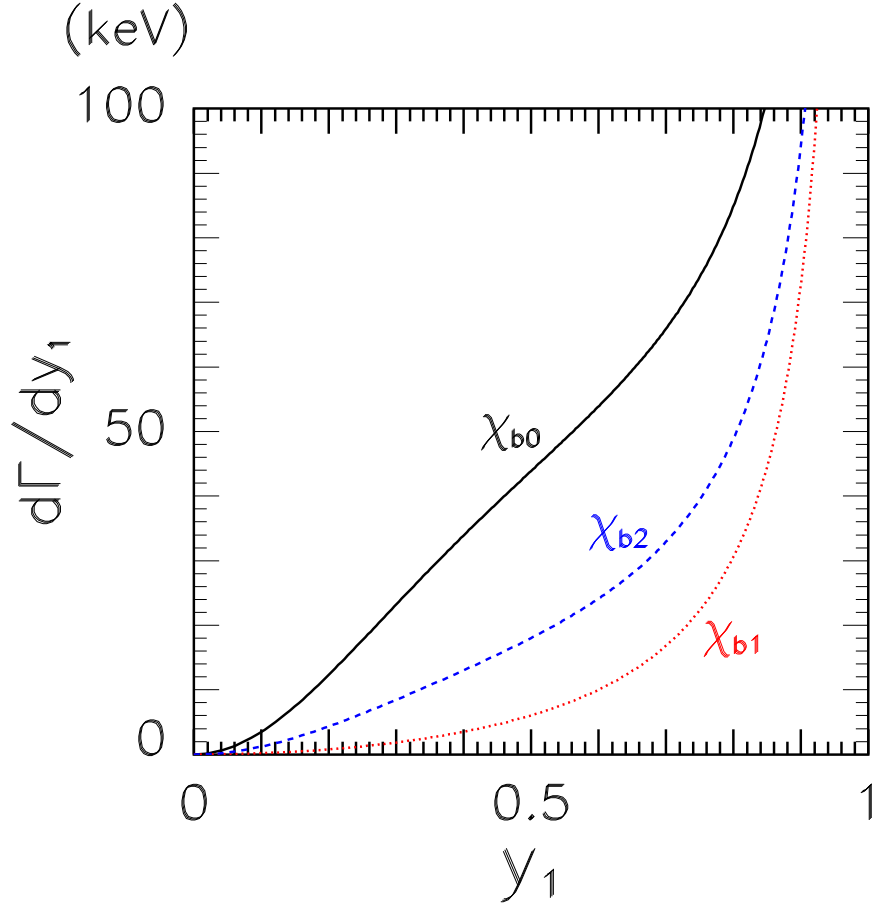}
\caption{
Distribution of the scaled momentum $y_1$ for the charm quark in decays of 
the $\chi_{bJ}$ for 
$J=0$ (solid line), 1 (dotted line), and 2 (dashed line)
for $\alpha_s=0.215$,
$\langle \mathcal{O}_1\rangle_{\chi_b}=2.03$ GeV$^5$,
and $m_b=4.6$ GeV.
}
\label{fig-dGamdx1}%
\end{figure}

The charm-quark momentum distributions in the decays of 
$\chi_{b0}$, $\chi_{b1}$, and $\chi_{b2}$ are illustrated in
Fig.~\ref{fig-dGamdx1}.  For the ratio $r$, which is defined in
Eq.~(\ref{r:c/b}), we choose the value $r = 4 m_D^2/m_{\chi_{bJ}}^2$,
which is equivalent in the nonrelativistic approximation
that we use in the calculation, but more correctly reflects the 
physical kinematics.
Here, $m_D$ is the average of the masses of the $D^0$ and $D^+$ 
and $m_{\chi_{bJ}}$ is the mass of the $\chi_{bJ}$ state.

If we use the most recent numerical values for the masses in 
Ref.~\cite{Yao:2006px}, we find that
this ratio is 0.1434, 0.1424, and 0.1419 for $J=0$, 1, and 2, 
respectively, for the $1P$ multiplet and
0.1331, 0.1325, and 0.1322 for $J=0$, 1, and 2, respectively, 
for the $2P$ multiplet.
For $y<1$, the color-octet terms in Eq.~(\ref{dGamdx1})
do not contribute at leading order in $\alpha_s$.
The normalizations of the momentum distributions for $y<1$
therefore depend only on the combination
$\alpha_s^3 \langle  \mathcal{O}_1 \rangle_{\chi_b}/m_b^4$.
We choose 
$\langle  \mathcal{O}_1 \rangle_{\chi_b} \approx 2.03 \ \textrm{GeV}^5$,
as is given in Eq.~(\ref{O11P}).  We take the bottom-quark mass 
to be the one-loop pole mass: 
$m_b = m_b^{\textrm{(pole)}} \approx 4.6$ GeV.
We take $\alpha_s$ to be the running coupling constant at the scale 
$m_b^{\textrm{(pole)}}$: $\alpha_s \approx 0.215$.
The $y_1$ distributions for $\chi_{b0}$, $\chi_{b1}$, and $\chi_{b2}$ 
in the $1P$ multiplet are shown  in Fig.~\ref{fig-dGamdx1}.
As is expected from our discussion in Sec.~\ref{color-singlet-coeffs},
all three curves diverge as $1/(1-y_1)$ 
as $y_1 \to 1$. There are also singular distributions with support 
only at $y_1 = 1$ that cannot be seen in the figure. 
As we have already mentioned,
the singular distributions are such that 
the integrals of the $y_1$ distributions
over an interval in $y_1$ that includes the endpoint
$y_1=1$ are finite.

\section{Total charm production rate}
\label{sec:rate}%
\subsection{Short-distance coefficients}

The inclusive charm production rate in decays of the $\chi_{bJ}$
can be calculated by integrating the differential rate in
Eq.~(\ref{dGamdx1}).
The integral of the color-octet coefficient 
in Eq.~(\ref{dA8cdx1}) is trivial:
\begin{eqnarray}
A_8^{(c)} = \frac{(1 + r/2)\sqrt{1-r}}{3} \pi \alpha_s^2.
\label{A8c}%
\end{eqnarray}
The required integrals for the color-singlet coefficients in 
Eq.~(\ref{dAJcdx1}) are tabulated in Appendix~\ref{app:intx1}.
These coefficients reduce to
\begin{subequations}
\begin{eqnarray}
A_0^{(c)}(\Lambda)
&=&
\frac{C_F \alpha_s^3}{N_c}
\left\{
\left[ \frac{2(2+r)}{9} \log \frac{8(1-r)m_b}{r\Lambda} - \frac{58+23r}{27}
	\right] \sqrt{1-r}
+ \frac{5}{9} \log \frac{1+\sqrt{1-r}}{1-\sqrt{1-r}} 
\right\},
\nonumber
\\
\\
A_1^{(c)}(\Lambda)
&=&
\frac{C_F \alpha_s^3}{N_c}
\left\{
\left[ \frac{2(2+r)}{9} \log \frac{8(1-r)m_b}{r\Lambda} - \frac{16+11r}{27}
	\right] \sqrt{1-r}
- \frac{4}{9} \log \frac{1+\sqrt{1-r}}{1-\sqrt{1-r}} 
\right\},
\label{A1:charm}%
\nonumber\\
\\
A_2^{(c)}(\Lambda)
&=&
\frac{C_F \alpha_s^3}{N_c}
\left\{
\left[ \frac{2(2+r)}{9} \log \frac{8(1-r)m_b}{r\Lambda} - \frac{116+91r}{135}
	\right] \sqrt{1-r}
- \frac{8}{45} \log \frac{1+\sqrt{1-r}}{1-\sqrt{1-r}} 
\right\}.
\nonumber
\\
\end{eqnarray}
\label{AJ:charm}%
\end{subequations}

\subsection{Comparison with previous results in the limit $\bm{m_c\to 0}$}

The limiting value of the color-octet coefficient $A_8^{(c)}$ 
in Eq.~(\ref{A8c}) as $r \to 0$ is $\frac{1}{3} \pi \alpha_s^2$, 
which agrees with the coefficient of $n_f$ in 
the leading-order result for $A_8$ in Eq.~(\ref{A8}).
The limiting behaviors of the color-singlet coefficients $A_J^{(c)}$
as $r \to 0$ are given by
\begin{subequations}
\begin{eqnarray}
A_0^{(c)}(\Lambda)
&\longrightarrow&
 \frac{C_F \alpha_s^3}{N_c}
\left( 
  \log \frac{4}{r} 
+ \frac{4}{9} \log \frac{2m_b}{\Lambda} - \frac{58}{27}
\right) ,
\\
A_1^{(c)}(\Lambda)
&\longrightarrow&
 \frac{C_F \alpha_s^3}{N_c}
\left( \frac{4}{9} \log \frac{2m_b}{\Lambda} - \frac{16}{27} \right) 
,
\label{A1charm:r=0}%
\\
A_2^{(c)}(\Lambda)
&\longrightarrow&
 \frac{C_F \alpha_s^3}{N_c}
\left( 
 \frac{4}{15} \log \frac{4}{r} 
+\frac{4}{9} \log \frac{2m_b}{\Lambda} - \frac{116}{135}
\right).
\end{eqnarray}
\label{AJcharm:r=0}%
\end{subequations}

The coefficients $A_0^{(c)}$ and $A_2^{(c)}$ in Eqs.~(\ref{AJcharm:r=0})
contain logarithms of $r$, and they therefore diverge in the limit $m_c \to 0$.
In the inclusive decay rates of the $\chi_{b0}$ and the $\chi_{b2}$,
the logarithmic sensitivity of the short-distance coefficients to $m_c$ 
is canceled by a correction to the decay rate for 
$b \bar b \to g g$ from virtual $c \bar c$ pairs.
The corrections of order $\alpha_s^3$ to
the $A_J$ from virtual charm quarks are given by
\begin{equation}
A_J^{(\textrm{virtual\ } c)}=-2i\Pi(0)A_J=
 \frac{2\alpha_s}{3\pi} A_J\log \frac{m_c}{\mu},
\end{equation}
where $\Pi(k^2)$ is the $\overline{\textrm{MS}}$-subtracted
quark-loop contribution to the gluon vacuum polarization at invariant 
four-momentum squared $k^2$, and the coefficients $A_J$ that are 
nonzero at order $\alpha_s^2$ are given in Eqs.~(\ref{AB:alphas2}).
Then we have
\begin{subequations}
\begin{eqnarray}
A_0^{(\textrm{virtual\ } c)} &=& 
  \frac{2C_F \alpha_s^3}{N_c}
\log\frac{m_c}{\mu},
\\
A_1^{(\textrm{virtual\ } c)} &=&  0,
\\
A_2^{(\textrm{virtual\ } c)} &=&  
  \frac{8C_F \alpha_s^3}{15N_c}
\log\frac{m_c}{\mu}.
\end{eqnarray}
\end{subequations}
where $\mu$ is the renormalization scale associated with
regularizing the ultraviolet divergence of the quark-loop
contributions to the gluon propagator. 
Upon adding these terms to the coefficients $A_J^{(c)}$ in 
Eqs.~(\ref{AJ:charm}), we see that
the logarithmic dependence on $m_c$ cancels
and we can take the limit $m_c \to 0$.  The sum of $A_J^{(c)}$
and $A_J^{(\textrm{virtual\ } c)}$ reduces in this limit to
\begin{subequations}
\begin{eqnarray}
\lim_{m_c\to 0} \left( A_0^{(c)}+A_0^{(\textrm{virtual\ } c)}
                \right)&=&
\frac{C_F \alpha_s^3}{N_c}
\left( 
  2 \log \frac{2m_b}{\mu}
+ \frac{4}{9} \log \frac{2m_b}{\Lambda} - \frac{58}{27}
\right) ,
                \\
\lim_{m_c\to 0} \left( A_1^{(c)}+A_1^{(\textrm{virtual\ } c)}
                \right)&=&
 \frac{C_F \alpha_s^3}{N_c}
\left( \frac{4}{9} \log \frac{2m_b}{\Lambda} - \frac{16}{27} \right),
                \\
\lim_{m_c\to 0} \left( A_2^{(c)}+A_2^{(\textrm{virtual\ } c)}
                \right)&=&                                
 \frac{C_F \alpha_s^3}{N_c}
\left( 
 \frac{8}{15} \log \frac{2m_b}{\mu} 
+\frac{4}{9} \log \frac{2m_b}{\Lambda} - \frac{116}{135}
\right).
\end{eqnarray}
\end{subequations}
These results agree with the coefficients of $n_f$ in the 
next-to-leading-order calculation of $A_J$ in Ref.~\cite{Petrelli:1997ge},
once one takes into account the different normalization convention 
for $\langle\mathcal{O}_1 \rangle_{\chi_b}$
that is used in Ref.~\cite{Petrelli:1997ge}.
\subsection{Fraction of charm decays}

\begin{figure}[htb]
\includegraphics*[width=12cm,angle=0,clip=true]{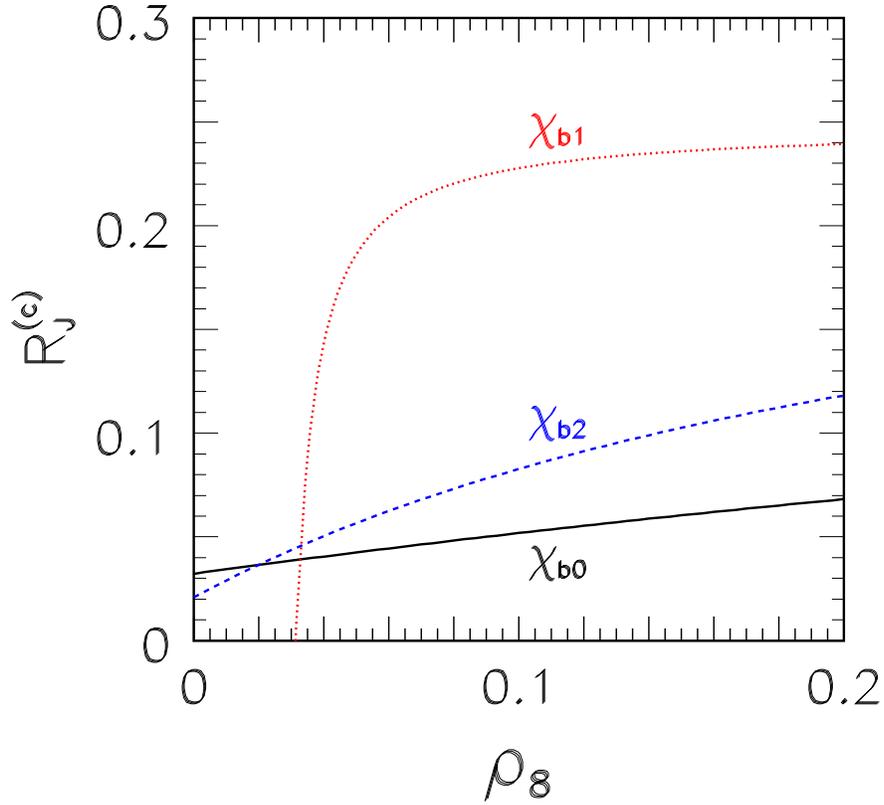}
\caption{Fractions $R^{(c)}_J$ of the annihilation decays of the $\chi_{bJ}$ 
that contain charm hadrons as functions of the ratio 
$\rho_8=m_b^2 \langle \mathcal{O}_8\rangle^{(m_b)}_{\chi_b}/%
\langle \mathcal{O}_1 \rangle_{\chi_b}$
for $J=0$ (solid line), 1 (dotted line), and 2 (dashed line).
}
\label{fig-RcJ}%
\end{figure}

The fraction $R^{(c)}_J$  of the decays of $\chi_{bJ}$
into light hadrons that include charm is given by the ratio 
of the NRQCD factorization formulas in Eqs.~(\ref{Gam-c}) and
(\ref{NRQCD:chi}):
\begin{equation}
R^{(c)}_J = 
\frac{A_J^{(c)}(m_b) \, \langle \mathcal{O}_1 \rangle_{\chi_b}
     + A_8^{(c)} \, m_b^2  \langle \mathcal{O}_8  \rangle^{(m_b)}_{\chi_b}}
       {A_J(m_b) \, \langle \mathcal{O}_1 \rangle_{\chi_b} 
     + A_8 \, m_b^2 \langle \mathcal{O}_8  \rangle^{(m_b)}_{\chi_b}} .
\label{RcJ}%
\end{equation}
The short-distance coefficients in the numerator are given at leading
order in $\alpha_s$ in Eqs.~(\ref{A8c}) and (\ref{AJ:charm}). The
short-distance coefficients in the denominator are given at leading
order in $\alpha_s$ in Eqs.~(\ref{AB:alphas2}) and (\ref{A1}).
In Fig.~\ref{fig-RcJ}, we show the fractions $R^{(c)}_J$
as a functions of the dimensionless ratio 
$\rho_8$ that is defined in Eq.~(\ref{rho8}).
These fractions  $R^{(c)}_J$ are sufficiently sensitive to
$\rho_8$ that $\rho_8$
could be determined phenomenologically from
measurements of the inclusive branching fractions of 
the $\chi_{bJ}$ into charm.

A simple physical constraint that can be imposed on the color-octet matrix 
element is that both the numerator and denominator in Eq.~(\ref{RcJ}) 
should be positive. If we use the leading-order approximations for the 
coefficients, then the strongest constraint comes from the positivity of 
the numerator for $J=1$.  This constraint requires that $\rho_8 > 0.032$.

\section{Charm-meson momentum distribution}
\label{sec:meson}%

In Sec.~\ref{sec:momentum}, we calculated the momentum distribution 
of the charm quark in decays of the $\chi_{b0}$, $\chi_{b1}$, and $\chi_{b2}$.
Once it is created, a charm quark will hadronize with nearly 100\% probability
into a charm hadron.  The charm hadron can be the $D^0$, $D^+$, $D_s$, 
or $\Lambda_c$, which are stable under strong and electromagnetic interactions,  
or it can be an excited charm hadron whose decay products include
the $D^0$, $D^+$, $D_s$, or $\Lambda_c$.   The effects of hadronization 
make the momentum spectrum of the charm hadron much softer than 
the momentum spectrum of the original charm quark. 

The fragmentation of a charm quark into a charm hadron can be studied 
by using $e^+ e^-$ collisions.  At leading order in $\alpha_s$, the production 
of a charm hadron in $e^+ e^-$ collisions with center-of-mass energy 
$\sqrt{s}$ proceeds through the creation of a $c$ and a $\bar c$ with 
momenta $\frac{1}{2} \sqrt{s - 4 m_c^2}$, followed by the fragmentation of 
the $c$ into the charm hadron.  Since the initial quark has a well-defined 
momentum, a measurement of the momentum distribution 
of the charm hadrons provides a measurement of the fragmentation process.
The CLEO and Belle Collaborations have measured the momentum 
distributions of various charm hadrons in $e^+ e^-$ annihilation
at center-of-mass energies near 10.6 GeV \cite{Artuso:2004pj,Seuster:2005tr}.  
This energy is fairly close to the masses of the $\chi_b$ states, 
which are near 9.9 GeV for the $1P$ multiplet and near 10.3 GeV for 
the $2P$ multiplet. The results of Refs.~\cite{Artuso:2004pj,Seuster:2005tr} 
show that the effects of hadronization are large.  It is convenient to 
describe them in terms  of the scaled momentum $y$ that is obtained by 
dividing the momentum by its maximum possible value.  At leading order in 
$\alpha_s$, the distribution for the charm quark is a Dirac delta 
function at $y=1$.  The peaks of the distributions in $y$ for the charm 
hadrons measured in Ref.~\cite{Seuster:2005tr} range from 0.59 to 0.68. 

A simple way to illustrate the effects of hadronization is to use a fragmentation 
approximation in which the  charm-hadron momentum distribution is given 
by the convolution of the momentum distribution of the charm quark with 
a fragmentation function.  The fragmentation function $D_{c \to D}(z)$
gives the probability distribution for a charm quark with plus component
of momentum $E_1 + \bm{p}_1$ to hadronize into a charm hadron $D$ 
with plus component of momentum $E_D + \bm{p}_D= z(E_1 + \bm{p}_1)$.
It is convenient to scale the plus component of momentum by its maximum 
possible value in the annihilation of $b \bar b$ at threshold.  The relation
between the scaled plus component $z_1$ and the scaled three-momentum 
$y_1$ of the charm quark is
\begin{equation}
z_1 = \frac{\sqrt{(1-r)y_1^2 + r} + \sqrt{1-r} \, y_1}{1 + \sqrt{1-r}}.
\label{ztoy}
\end{equation}
The inverse relation is
\begin{equation}
 y_1 = \frac{ (1 + \sqrt{1-r})^2 z_1^2 - r}{2 \sqrt{1-r} (1 + \sqrt{1-r}) z_1}.
\label{ytoz}
\end{equation}
If we neglect the difference between the mass of the quark and the 
mass of the charm hadron, there are similar relations between the 
scaled components $z_D$ and $y_D$ of the four-momentum of the 
charm hadron.  The fragmentation approximation for the momentum 
distribution of the charm hadron can then be written as
\begin{eqnarray}
\frac{d\Gamma}{d y_D} 
&=&
\frac{dz_D}{dy_D} \int_{z_D}^{1} \frac{dz_1}{z_1} \,
D(z_D/z_1)\, \frac{dy_1}{dz_1}
\frac{d\Gamma}{d y_1} 
\nonumber\\
&=&
\frac{\sqrt{1-r}}{\sqrt{(1-r)y_D^2+r }}
\int_{y_D}^1 dy_1
  \mathcal{D}\left(
  \frac{\sqrt{(1-r)y_D^2+r} +\sqrt{1-r}y_D }
       {\sqrt{(1-r)y_1^2+r} +\sqrt{1-r}y_1 }
\right) \frac{d\Gamma}{dy_1},
\label{dGamD}%
\end{eqnarray}
where $\mathcal{D}(z) = z D(z)$.
The expression for $d\Gamma/dy_D$ in Eq.~(\ref{dGamD}), when integrated
over $y_D$, does not preserve the normalization of the total cross
section $\int(d\Gamma/dy_1)\,dy_1$,
unless one takes the approximation of neglecting
$m_c$ in comparison to $m_b$, {\it i.e.} setting $r=0$, in the relations
(\ref{ztoy}) and (\ref{ytoz}) and in the limits of integration.
In this approximation, $z_1=y_1$ and
$z_D=y_D$. The change in the normalization of the total cross section is
negative and is of order $r$. This change is at the level of the error
in the fragmentation approximation itself, which is derived from QCD by
neglecting corrections on the order of the square of the quark mass 
divided by the hard-scattering momentum \cite{Collins:1989gx}.

The Belle Collaboration determined optimal values of the parameters for 
analytic parameterizations of the fragmentation functions for various 
charm hadrons by comparing their measured momentum distributions  
with the distributions predicted by Monte Carlo generators and 
fragmentation functions \cite{Seuster:2005tr}. The best fits were obtained 
by using fragmentation functions that are functions of $z$ and the 
transverse momentum $p_\perp$.  Of the fragmentation functions that are 
functions of $z$ only, the best fit was usually obtained by using 
the very simple Kartvelishvili fragmentation function:
\begin{equation}
D_{c  \to  D}(z)= N_D z^{\alpha_D} (1-z).
\label{D-Kart}%
\end{equation}
The fit for the $D^+$ was better than that for the 
$D^0$, presumably because 
the momentum distribution for the $D^+$ has smaller contributions from the 
feeddown from decays of the $D^{*0}$ and the $D^{*+}$. 
For the $D^+$, the best fit 
for the exponent in Eq.~(\ref{D-Kart}) was $\alpha_{D^+} = 4$.  
The resulting
 fragmentation function has a peak at $z = 0.8$.  The integral 
$\int_0^1 dz D_{c \to  D}(z)$ is the fragmentation probability for the 
charm hadron $D$.  From Table~X of Ref.~\cite{Seuster:2005tr}, we can 
infer that the inclusive fragmentation probability for the $D^+$, including 
the feeddown from decays of the $D^{*+}$, is approximately 0.268.  
This fixes the normalization factor in Eq.~(\ref{D-Kart}) 
to be $N_{D^+} = 8.04$.

\begin{figure}
\centerline{\includegraphics*[width=12cm,angle=0,clip=true]{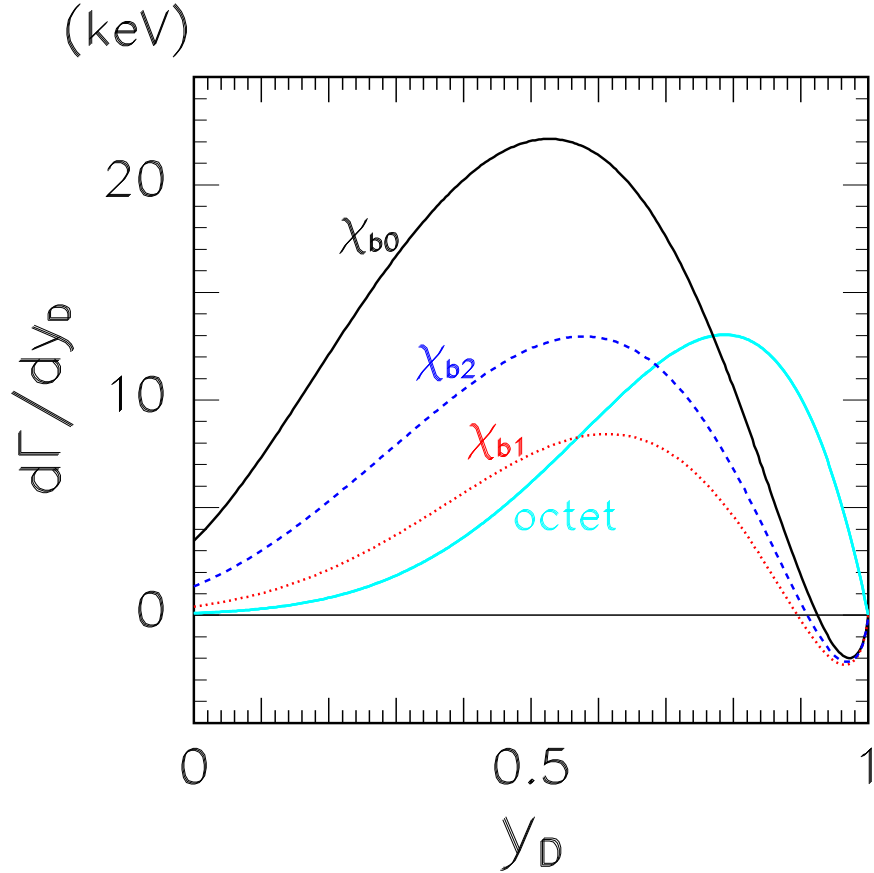}}
\caption{
Distribution of the scaled momentum $y_D$ for the charm meson $D^+$ 
in decays of the $\chi_{bJ}$ for 
$J=0$ (solid line), 1 (dotted line), and 2 (dashed line)
for $\rho_8=0.1$.
Also shown is the color-octet contribution to the
distributions (light solid line),
which is the same, to within about $2~\%$, for $J=0$, 1, and 2
and is already included in the other three curves.
The unphysical negative behavior of the color-singlet contributions 
near the endpoint at $y_D=1$ might be eliminated by resumming 
logarithmic corrections to all orders, as is described in the text.
}
\label{fig-dGamdyD}%
\end{figure}

The fragmentation approximation to the $D^+$ momentum distribution 
that is given by Eq.~(\ref{dGamD}) is shown in Fig.~\ref{fig-dGamdyD},
where the 
fragmentation function is given in Eq.~(\ref{D-Kart}).
We have set  $r = 4m_D^2/m_{\chi_{bJ}}^2$, 
$\alpha_s=\alpha_s(4.6~\textrm{GeV}) = 0.215$, and
$\langle  \mathcal{O}_1 \rangle_{\chi_b} \approx 2.03 \ \textrm{GeV}^5$,
as is given in Eq.~(\ref{O11P}), and we have chosen $\rho_8=0.1$.  
Within the fragmentation 
approximation, the peaks in the momentum distributions of 
the $D^+$ from decays of the $\chi_{bJ}$
are at $y_D = 0.53$, 0.61 and 0.58 for $J=0$, 1 and 2, respectively.
Also  shown in Fig.~\ref{fig-dGamdyD} is the color-octet contribution to 
all three distributions, which peaks at $y_D = 0.79$.
As we have mentioned above, the normalization of the total cross section
decreases in the fragmentation approximation by an amount of order $r$.
In the present case, the fragmentation approximation decreases the total
cross sections by about $2.6\%$, $1.0\%$, and $1.9\%$ for $J=0$, $1$, 
and $2$, respectively.

The momentum distributions in Fig.~\ref{fig-dGamdyD} have unphysical
negative values near the endpoint at $y_D=1$. The momentum distributions
very near the endpoint are dominated by the $[1/(1-x)]_{\sqrt{r}}$ terms
in the coefficients $d A_J^{(c)}(\Lambda)/dx_1$ in Eqs.~(\ref{dAJcdx1}).
 If we use Eq.~(\ref{sing-xy}) to transform the distribution in $x_1$
into a distribution in $y_1$, then the negative terms come from the last 
term in the definition of $[1/(1-y)]_+$ in Eq.~(\ref{plus-def}). 
The limiting behavior as $y_D \to 1$ from this term in the momentum 
distribution is
\begin{eqnarray}
\frac{d\Gamma}{d y_D} \sim
- D(z_D) \log \left( \frac{1}{1-y_D} \right) 
\frac{2(2+r)\sqrt{1-r} C_F \alpha_s^3 
      \langle \mathcal{O}_1 \rangle_{\chi_b}}{9 N_c m_b^4}.
\label{dGam-lim}%
\end{eqnarray}
If we use the Kartvelishvili fragmentation function [Eq.~(\ref{D-Kart})], 
then all other terms in $d\Gamma/dy_D$ vanish
linearly in $1-y_D$ as $y_D \to 1$. The fragmentation function $D(z_D)$
in Eq.~(\ref{dGam-lim}) also vanishes linearly in $1-y_D$ as $y_D \to
1$, but the logarithm  approaches $- \infty$, and so this negative term
dominates sufficiently close to the end point.

As we mentioned with regard to the distributions in $x_1$ in
Sec.~\ref{color-singlet-coeffs}, such unphysical contributions arise
because, near $y_D=1$, large logarithms of $1-x_1$ cause the
perturbation expansion in $\alpha_s$ to break down. We expect that
resummation of these logarithms to all orders in perturbation theory
would cure the distribution in $y_D$ of these unphysical effects.

\section{Summary}

We have used the NRQCD factorization formalism to calculate the
inclusive decay rate of the spin-triplet bottomonium states $\chi_{bJ}$
into charm hadrons. In Eq.~(\ref{Gam-c}), the decay rates are
expressed in terms of two independent nonperturbative factors for each
$P$-wave multiplet: $\langle \mathcal{O}_1 \rangle_{\chi_b}$ and $\langle
\mathcal{O}_8 \rangle ^{(\Lambda)}_{\chi_b}$. The coefficients of these
factors were calculated to leading order in $\alpha_s$ using
perturbative matching. Our results for the coefficients that are
differential in the $c$-quark energy fraction are given in
Eqs.~(\ref{dA8cdx1}) and (\ref{dAJcdx1}). Our results for the
coefficients integrated over the $c$-quark energy fraction are given in
Eqs.~(\ref{A8c}) and (\ref{AJ:charm}). The ratios $R^{(c)}_J$ of the
decay rate of the $\chi_{bJ}$ into light hadrons that include charm
and the decay rate into all light hadrons are shown in
Fig.~\ref{fig-RcJ} as a function of the ratio $\rho_8$ of the NRQCD matrix
elements.  The ratios $R^{(c)}_J$ are sufficiently sensitive to $\rho_8$
that measurements of the branching fraction  of the $\chi_{bJ}$ into
charm could be used to make a phenomenological determination of 
the $\langle \mathcal{O}_8 \rangle ^{(\Lambda)}_{\chi_b}$. 
These matrix elements could then be used to
predict the partial widths into light hadrons for  all four states in
the $P$-wave bottomonium multiplet.

We also calculated the momentum distribution of the charm quark from the
decays of the $\chi_{bJ}$. 
We obtained a simple approximation to the momentum
distribution for charm mesons in $\chi_{bJ}$ decay by convolving the
charm-quark momentum distribution with a fragmentation function for $c
\to D$ that was measured in $e^+ e^-$ collisions. The
charm-meson momentum distributions for the $\chi_{bJ}$ are shown in
Fig.~\ref{fig-dGamdyD} as functions of the scaled momentum variable
$y_D$ for $\rho_8=0.1$. The CLEO-III experiment and the $B$ factory
experiments may be able to measure the momentum distributions of charm
hadrons in $\chi_{bJ}$ decay.  One unsatisfactory aspect of the
theoretical momentum distributions in  Fig.~\ref{fig-dGamdyD} is the
unphysical negative behavior of the distributions near the endpoint at
$y_D=1$. 
We expect that this difficulty could be overcome by resumming logarithmic
corrections to all orders in $\alpha_s$. 
The region near the endpoint also receives
large contributions that are formally of higher order in the NRQCD
velocity expansion. Such contributions can be resummed to all orders in
$v$ by making use of a shape function. The completion of these
resummation calculations would allow one to make quantitative
comparisons between theoretical predictions and experimental
measurements of the momentum distributions of the charm hadrons that are
produced in $\chi_{bJ}$ decays.

\begin{acknowledgments}
%
We thank Roy Briere for suggesting this problem and for useful discussions.
E.~Braaten thanks KITP for its hospitality while this work was being completed.
J.~Lee thanks the High Energy Physics Theory Group at Argonne National 
Laboratory for its hospitality while this work was carried out.
Work by G.~T.~Bodwin in the High Energy Physics Division at Argonne
National Laboratory is supported by the U.~S.~Department of Energy,
Division of High Energy Physics, under Contract No.~DE-AC02-06CH11357.
The work of E.~Braaten was supported in part by the U.~S.~Department 
of Energy, Division of High Energy Physics, under grant 
No.~DE-FG02-91-ER40690. 
The work of D.~Kang was supported by the Korea Research Foundation
under grant KRF-2006-612-C00003.
The work of J.~Lee was supported by the Korea Research Foundation
under MOEHRD Basic Research Promotion grant KRF-2006-311-C00020
and by the Basic Research Program of the Korea Science and Engineering 
Foundation (KOSEF) under grant No.~R01-2005-000-10089-0.

\end{acknowledgments}

\appendix

\section{Dimensionally Regularized three-Body Phase Space
\label{sec:dphi3}%
}

In $d=4-2\epsilon$ space-time dimensions, the three-body phase space 
is defined by
\begin{equation}
d\Phi_3=(2\pi)^{d}\delta^{(d)}(P-p_1-p_2-p_3)
\frac{d^{d-1}{p}_1}{(2\pi)^{d-1}2E_1}
\frac{d^{d-1}{p}_2}{(2\pi)^{d-1}2E_2}
\frac{d^{d-1}{p}_3}{(2\pi)^{d-1}2E_3},
\label{PS3-def-app}%
\end{equation}
where $E_i$ and $p_i$ are the energy and four-momentum of the particle 
$i$ in the final state with mass $m_i$, and $P=p_1+p_2+p_3$.
We evaluate $d\Phi_3$ in
the center-of-momentum frame, $P=(\sqrt{P^2},\bm{0})$, 
where the resulting expressions are most compact.
In any decay with a three-body final state, the
squared matrix element, summed over spin states, is a Lorentz scalar,
depending only on the four momenta $P$, $p_1$, $p_2$, and $p_3$.
By using energy-momentum conservation, it can be seen that
all possible scalar products of momenta can be expressed in terms
of $E_i$'s. Therefore the spin-summed matrix element squared depends 
only on the energies $E_i$.

Integrating out $\bm{p}_2$ and all angles except for the relative  
angle between $\bm{p}_1$ and $\bm{p}_3$, we obtain
\begin{equation}
d\Phi_3
=
\frac{(4\pi)^{2\epsilon}\,\Gamma\left(\frac{3}{2}\right)}
     {\Gamma(1-\epsilon)\Gamma\left(\frac{3}{2}-\epsilon\right)}
\frac{ (|\bm{p}_1||\bm{p}_3|)^{d-3} }{E_2}
\sin^{d-4}\theta_{13}\,
\delta(\sqrt{P^2}-E_1-E_2-E_3) 
\frac{dE_1 dE_3}{32\pi^3}d\cos\theta_{13},
\label{PS3-1-app}%
\end{equation}
where $\theta_{13}$ is the angle between $\bm{p}_1$ and $\bm{p}_3$
in the center-of-momentum frame. The angle $\theta_{13}$ is fixed by the energy 
delta function:
\begin{equation}
E_2=\sqrt{|\bm{p}_1|^2+|\bm{p}_3|^2
+2|\bm{p}_1||\bm{p}_3|\cos\theta_{13}+m_2^2}.
\label{e2-cos13}%
\end{equation}
By solving Eq.~(\ref{e2-cos13}) in the center-of-momentum frame, 
we can express $\sin^2\theta_{13}$ in terms of the magnitudes of 
three-momenta for the final-state particles:
\begin{equation}
\sin^2 \theta_{13} =
-\lambda(\bm{p}_1^2,\bm{p}_2^2,\bm{p}_3^2)/(4 \bm{p}_1^2 \bm{p}_3^2),
\label{sin13-app}%
\end{equation}
where $\lambda(x,y,z) = x^2+y^2+z^2 -2(xy+yz+zx)$.
The physical region can be determined from the expression
\begin{equation}
-\lambda(\bm{p}_1^2,\bm{p}_2^2,\bm{p}_3^2)
=
(|\bm{p}_1|+|\bm{p}_2|+|\bm{p}_3|)
(|\bm{p}_1|+|\bm{p}_2|-|\bm{p}_3|)
(|\bm{p}_2|+|\bm{p}_3|-|\bm{p}_1|)
(|\bm{p}_3|+|\bm{p}_1|-|\bm{p}_2|)>0.
\label{lam-physical}%
\end{equation}
Substituting Eq.~(\ref{sin13-app}) into Eq.~(\ref{PS3-1-app}) and
changing the integration variable from $\cos\theta_{13}$ to $E_2$,
using Eq.~(\ref{e2-cos13}), we obtain
\begin{equation}
d\Phi_3 =
\frac{(4\pi)^{2\epsilon}}{\Gamma(2-2\epsilon)}
\delta(\sqrt{P^2}-E_1-E_2-E_3) 
\frac{dE_1 dE_2 dE_3}
     {32 \pi^3\left[ -\lambda(\bm{p}_1^2,\bm{p}_2^2,\bm{p}_3^2) 
\right]^{\epsilon}}.
\end{equation}

In the calculations in this paper, we study the case $p_1^2=p_2^2=m_c^2$, 
$p_3^2=0$, and $\sqrt{P^2}=2E_b$.
We express the energy variables in terms of dimensionless variables
$x_i=E_i/E_b$. Then 
\begin{equation}
\cos\theta_{13}
=
\frac{2E_b(E_b-E_1-E_3)+E_1E_3}
{|\bm{p}_1|E_3}
=
\frac{(1-x_1)/x_3 - a(x_1)}{b(x_1)},
\label{cos13}%
\end{equation}
where $a(x_1)$ and $b(x_1)$ are defined by
\begin{subequations}
\begin{eqnarray}
a(x_1)&=&1-\tfrac{1}{2}x_1,
\\
b(x_1)&=&\tfrac{1}{2}\sqrt{x_1^2-r}.
\end{eqnarray}
\label{ab-def}%
\end{subequations}
The ranges of integrals are determined from Eq.~(\ref{lam-physical}):
\begin{subequations}
\begin{eqnarray}
\label{x3-range-app}%
\sqrt{r}\leq &x_1& \leq1
,
\\
\label{x1-range-app}%
x_3^- \leq &x_3& \leq x_3^+
,
\end{eqnarray}
\label{range-app}%
\end{subequations}
where $x_3^\pm$ are defined by
\begin{equation}
x_3^\pm 
=\frac{1-x_1}{a(x_1)\mp b(x_1)}.
\label{ab-app}%
\end{equation}

\section{Evaluation of Integrals $\bm{I_n(x_1)}$}
\label{sec:In}%

In this appendix, we evaluate the integrals
$I_n(x_1)$ that are defined in Eq.~(\ref{In}):
\begin{equation}
I_n(x_1)=\int_{x_3^-}^{x_3^+} \frac{(1-x_1)^n dx_3}
          {x_3^{n+2+2\epsilon}(1-\cos^2\theta_{13})^\epsilon}.
\label{In-0}%
\end{equation}
where the bounds $x_3^\pm$ of the integral are given in 
Eqs.~(\ref{ab-def}) and (\ref{ab-app}).  The cosine of 
the angle $\theta_{13}$ is expressed as a function of $x_1$ and $x_3$
in Eq.~(\ref{cos13}).  By making the changes of variables
\begin{equation}
t=\frac{1-x_1}{x_3}=a(x_1)+b(x_1)\cos\theta_{13}
\label{tx3}%
\end{equation}
and using the relation for $\cos\theta_{13}$ in Eq.~(\ref{cos13}),
we can parametrize the $I_n(x_1)$ as
\begin{eqnarray}
I_n(x_1)&=&\frac{1}{(1-x_1)^{1+2\epsilon}}
          \int_{a(x_1)-b(x_1)}^{a(x_1)+b(x_1)} 
          \frac{t^{n+2\epsilon}dt}
          {(1-\cos^2\theta_{13})^\epsilon}
\nonumber\\
&=&
\frac{b(x_1)}{(1-x_1)^{1+2\epsilon}}
          \int_{-1}^{1}
          \frac{[a(x_1)+b(x_1) x]^{n+2\epsilon} dx}
          {(1-x^2)^\epsilon}.
\label{In-1}%
\end{eqnarray}
In the second line of Eq.~(\ref{In-1}), we used Eq.~(\ref{cos13}).
It is evident that the $t$ or $x$ integrals in Eq.~(\ref{In-1}) are
finite. The divergent part is contained in 
the factor $1/(1-x_1)^{1+2\epsilon}$,
which is manifestly logarithmically divergent in the infrared limit 
$x_1\to 1$. The integral of that factor 
over $x_1$ is proportional to 
$-1/(2\epsilon)$. The evaluation of the integrals $I_n(x_1)$, keeping the 
full $\epsilon$ dependence, is quite involved. 
However, in order to compute the pole in $\epsilon$ and the finite
term, we need only to expand
the coefficient of $1/(1-x_1)^{1+2\epsilon}$ in Eq.~(\ref{In-1}) 
to order $\epsilon$:
\begin{eqnarray}
I_n(x_1)&=&\frac{1}{(1-x_1)^{1+2\epsilon}}
\bigg\{
          \int_{a(x_1)-b(x_1)}^{a(x_1)+b(x_1)}
          {t^{n+2\epsilon}dt}
\nonumber\\
&&
-\epsilon\,
b(x_1)
 \int_{-1}^{1}
          [a(x_1)+b(x_1) x]^{n}\log(1-x^2) dx
\bigg\}+O(\epsilon)
\nonumber\\
&=&
\frac{1}{(1-x_1)^{1+2\epsilon}}
\bigg\{
\frac{ [a(x_1)+b(x_1)]^{n+1+2\epsilon}-[a(x_1)-b(x_1)]^{n+1+2\epsilon}
     }{n+1+2\epsilon}
-2\epsilon\, i_n(x_1)
\bigg\}
\nonumber\\[1ex]
&&+O(\epsilon),
\label{In-2}%
\end{eqnarray}
where the $i_n(x_1)$ for $n=$ 0, 1, and 2 are given by
\begin{subequations}
\begin{eqnarray}
i_0(x_1)&=&-2b(x_1)(1-\log 2),
\\
i_1(x_1)&=&-2a(x_1)b(x_1)(1-\log 2),
\\
i_2(x_1)&=&-2b(x_1)\left\{
\tfrac{1}{9}[b(x_1)]^2(4- 3\log 2)+[a(x_1)]^2(1-\log 2)\right\}.
\end{eqnarray}
\label{table-In}%
\end{subequations}

It is convenient to rewrite the divergent integral as a linear combination
of finite integrals and a singular integral involving a delta function.
As we have noted, all the factors except for $1/(1-x_1)^{1+2\epsilon}$
are regular functions of $x_1$. We denote the factor that
is the coefficient of $1/(1-x_1)^{1+2\epsilon}$ by $f(x_1)$. 
Therefore, we wish to study the integral
\begin{equation}
I=\int_{\sqrt{r}}^1 \frac{f(x_1)dx_1}{(1-x_1)^{1+2\epsilon}},
\label{I1}%
\end{equation}
where $f(x_1)$ is regular for any $x_1\in [\sqrt{r},1]$.
One can separate the divergent contributions to the integral
from the finite piece as follows.
\begin{equation}
I=
\int_{\sqrt{r}}^1 \frac{f(x_1)-f(1)}{(1-x_1)^{1+2\epsilon}} \,dx_1
+
f(1)\int_{\sqrt{r}}^1 \frac{dx_1}{(1-x_1)^{1+2\epsilon}}.
\label{I2}%
\end{equation}
The first term on the right side of Eq.~(\ref{I2})
is finite in the limit $\epsilon\to 0$. 
The second term on the right side of Eq.~(\ref{I2})
is singular in the limit $\epsilon\to 0$:
\begin{equation}
\int_{\sqrt{r}}^1 \frac{dx_1}{(1-x_1)^{1+2\epsilon}}=
-\frac{1}{2\epsilon(1-\sqrt{r})^{2\epsilon}}.
\label{I3}%
\end{equation}
Substituting Eq.~(\ref{I3}) into Eq.~(\ref{I2}), we obtain
\begin{equation}
I=
\int_{\sqrt{r}}^1 dx_1
\left[
\frac{f(x_1)-f(1)}{(1-x_1)^{1+2\epsilon}}
-\frac{f(x_1)\delta(1-x_1)}{2\epsilon(1-\sqrt{r})^{2\epsilon}}
\right].
\label{I4}%
\end{equation}
Expanding the $\epsilon$ dependence in the first term in 
Eq.~(\ref{I4}), we have
\begin{equation}
\frac{1}{(1-x_1)^{1+2\epsilon}}
=-\frac{\delta(1-x_1)}{2\epsilon(1-\sqrt{r})^{2\epsilon}}
+
\left[ \frac{1}{1-x_1} \right]_{\sqrt{r}}
-2\epsilon
\left[
\frac{\log(1-x_1)}{1-x_1}
\right]_{\sqrt{r}}
+O(\epsilon^2),
\label{I5}%
\end{equation}
where the distribution $[1/(1-x_1)]_{\sqrt{r}}$ 
is defined by Eq.~(\ref{root-def}).
Retaining the first two terms in  Eq.~(\ref{I5}), we obtain
\begin{subequations}
\begin{eqnarray}
\frac{I_0(x_1)}{(x_1^2-r)^\epsilon}&=&
 \left[
 \left(-\frac{1}{2\epsilon}+\log 2
 \right)\sqrt{1-r}
 +L_0(r)
 \right]\delta(1-x_1)
+
 \left[\frac{1}{1-x_1}\right]_{\sqrt{r}}\sqrt{x_1^2-r}
+O(\epsilon),
\nonumber\\
\\
\frac{I_1(x_1)}{(x_1^2-r)^\epsilon}&=&
 \left[
 \left(-\frac{1}{4\epsilon}-\frac{1}{4}+\frac{1}{2}\log 2
 \right)
\sqrt{1-r}
 +L_1(r)
 \right]\delta(1-x_1)
\nonumber\\
&&+
 \left\{\frac{1}{2}+\frac{1}{2}\left[\frac{1}{1-x_1}\right]_{\sqrt{r}}
 \right\}\sqrt{x_1^2-r}
+O(\epsilon),
\\
\frac{I_2(x_1)}{(x_1^2-r)^\epsilon}&=&
 \left\{
 \left[
 \left(-\frac{1}{24\epsilon}+\frac{1}{12}\log 2
 \right)(4-r)
 -\frac{1}{4}+\frac{r}{12}
 \right]\sqrt{1-r}
 +L_2(r)
 \right\}\delta(1-x_1)
\nonumber\\
&& +
 \left\{\frac{2-x_1}{3}
       +\frac{4-r}{12}\left[\frac{1}{1-x_1}
                            \right]_{\sqrt{r}} 
 \right\}\sqrt{x_1^2-r}
+O(\epsilon).
\end{eqnarray}
\label{In_int}%
\end{subequations}

The functions $L_n(r)$ are given by
\begin{eqnarray}
L_n(r)&=&
-\frac{1}{n+1} \left[(a+b)^{n+1} \log(a+b)-(a-b)^{n+1} \log(a-b) \right]
\nonumber\\
&&
+\frac{1}{2(n+1)} \left[(a+b)^{n+1} -(a-b)^{n+1} \right]
\log \left[ (1-\sqrt{r})^2 (1-r) \right],
\label{Ln}%
\end{eqnarray}
where $a\pm b=a(1)\pm b(1)=\tfrac{1}{2}(1\pm \sqrt{1-r})$.
These functions vanish in the massless limit $r\to 0$: $L_n(0)=0$.
The explicit expressions for $L_0(r)$, $L_1(r)$, and $L_2(r)$ are
\begin{subequations}
\begin{eqnarray}
L_0(r)&=&\frac{\sqrt{1-r}}{2}
          \log \frac{4(1-\sqrt{r})^2 (1-r)}{r}
-\frac{1}{2}\log\frac{1+\sqrt{1-r}}{1-\sqrt{1-r}}
,
\\
L_1(r)&=& \frac{\sqrt{1-r}}{4}
          \log \frac{4(1-\sqrt{r})^2 (1-r)}{r}
-\frac{2-r}{8}\log\frac{1+\sqrt{1-r}}{1-\sqrt{1-r}}
,
\\
L_2(r)&=&\frac{(4-r)\sqrt{1-r}}{24}
          \log \frac{4(1-\sqrt{r})^2 (1-r)}{r}
-\frac{4-3r}{24}\log\frac{1+\sqrt{1-r}}{1-\sqrt{1-r}}
.
\hspace{1.0cm}
\end{eqnarray}
\end{subequations}
\section{Evaluation of integrals over $\bm{x_3}$}
\label{app:intx3}%

In this appendix, we report the results of carrying out the integrations
over $x_3$ in the components of the $b\bar{b}$ differential widths
in Eqs.~(\ref{div}) and (\ref{Gamhat-fin}).

The integrals $I_n(x_1)$ that appear in the infrared-divergent 
functions $\hat{\Gamma}_{\textrm{div}}^J(x_1)$ in Eqs.~(\ref{div})
are defined by integrals over $x_3$ that are evaluated in 
Appendix~\ref{sec:In}.  Making use of these results, we
find that
\begin{subequations}
\begin{eqnarray}
\hat{\Gamma}_{\textrm{div}}^0(x_1)
 &=&
\left\{
\left(
\frac{2(2+r)}{9}
\left[ -\frac{1}{2\epsilon}+\log 2\right]
+\frac{5+r}{27}
\right)\sqrt{1-r}
\right.
\nonumber\\
&&
\hspace{2cm} \left.
 +\frac{2+r}{3} L_0(r) -\frac{4}{3}[L_1(r)-L_2(r) ]
\right\} \delta(1-x_1)
\nonumber\\
&&
+
\frac{2}{9}
\left\{1-2 x_1
     +(2+r)
      \left[\frac{1}{1-x_1}\right]_{\sqrt{r}}
\right\}
\sqrt{x_1^2-r}
,\\
\hat{\Gamma}_{\textrm{div}}^1(x_1)
&=&
\left\{
\left(
\frac{2(2+r)}{9}
\left[ -\frac{1}{2\epsilon}+\log 2\right]
+\frac{7-4 r}{54}
\right)\sqrt{1-r}
\right.
\nonumber\\
&&
\hspace{2cm} \left.
+ \frac{2+r}{6} L_0(r) +\frac{2}{3}[L_1(r)-L_2(r) ]
\right\} \delta(1-x_1)
\nonumber\\
&&
+
\frac{2}{9}
\left\{-\frac{1-2 x_1}{2}
     +(2+r)
      \left[\frac{1}{1-x_1}\right]_{\sqrt{r}}
\right\}
\sqrt{x_1^2-r}
,\\
\hat{\Gamma}_{\textrm{div}}^2(x_1)
&=&
\left\{
\left(
\frac{2(2+r)}{9}
\left[ -\frac{1}{2\epsilon}+\log 2\right]
+\frac{41-8 r}{270}
\right)\sqrt{1-r}
\right.
\nonumber\\
&&
\hspace{2cm} \left.
+\frac{7(2+r)}{30} L_0(r) -\frac{2}{15}[L_1(r)-L_2(r) ]
\right\}\delta(1-x_1)
\nonumber\\
&&
+
\frac{2}{9}
\left\{
         \frac{1-2 x_1}{10}
         +(2+r)\left[\frac{1}{1-x_1}\right]_{\sqrt{r}}
\right\}
\sqrt{x_1^2-r}
,
\end{eqnarray}
\label{div-eval}%
\end{subequations}
where the distribution $[1/(1-x_1)]_{\sqrt{r}}$ is defined in
Eq.~(\ref{plus-def}).

The infrared-finite functions $\hat{\Gamma}_{\textrm{fin}}^J(x_1)$
in Eqs.~(\ref{Gamhat-fin}) are defined by integrals over $x_3$ that are 
straightforward to evaluate.
The results of carrying out these integrations are
\begin{subequations}
\begin{eqnarray}
\hat{\Gamma}_{\textrm{fin}}^0(x_1) &=&
 \frac{2\sqrt{x_1^2-r}}{3}(9-11x_1)
+\left[
\frac{3}{2}+r-3x_1(1-x_1)
\right]
\log \frac{x_1+\sqrt{x_1^2-r}}{x_1-\sqrt{x_1^2-r}}
\nonumber\\&&
-\frac{1}{6}
\log\frac{2-x_1+\sqrt{x_1^2-r}}{2-x_1-\sqrt{x_1^2-r}},
\\
\hat{\Gamma}_{\textrm{fin}}^1(x_1)&=&
 \frac{x_1\sqrt{x_1^2-r}}{3}
-\frac{1}{3}
\log\frac{2-x_1+\sqrt{x_1^2-r}}{2-x_1-\sqrt{x_1^2-r}},
\\
\hat{\Gamma}_{\textrm{fin}}^2(x_1) &=&
\frac{\sqrt{x_1^2-r}}{15}(24-29 x_1)
+\frac{2}{5}\big[1+r-2x_1(1-x_1)\big]
\log \frac{x_1+\sqrt{x_1^2-r}}{x_1-\sqrt{x_1^2-r}}
  \nonumber\\
  &&
-\frac{1}{15}
\log\frac{2-x_1+\sqrt{x_1^2-r}}{2-x_1-\sqrt{x_1^2-r}}.
\end{eqnarray}
\label{fin-eval}%
\end{subequations}

\section{Evaluation of Integrals over $\bm{x_1}$\label{app:intx1}%
}

In this appendix, we tabulate the integrals that are required to obtain
the inclusive short-distance coefficients $A_{J}^{(c)}$ in Eq.~(\ref{AJ:charm}) from
the short-distance coefficients $dA_{J}^{(c)}$ in Eq.~(\ref{dAJcdx1})
that are differential in $x_1$.

Some of the basic integrals over $x_1$ are
\begin{subequations}
\begin{eqnarray}
\int_{\sqrt{r}}^{1} dx_1 \,
x_1  \sqrt{x_1^2-r}
&=&
\frac{1}{3}(1-r)^{3/2}
,\\
\int_{\sqrt{r}}^{1} dx_1 \,\,\,
\sqrt{x_1^2-r}
&=&
  \frac{\sqrt{1-r}}{2}
 -\frac{r}{4}
   \log \frac{1+\sqrt{1-r}}{1-\sqrt{1-r}}
,\\
\int_{\sqrt{r}}^{1} dx_1
\sqrt{x_1^2-r}
\left[\frac{1}{1-x_1}\right]_{\sqrt{r}}
&=&
-\sqrt{1-r}\left[1+\frac{1}{2}\log\frac{r}{4}-\log(1+\sqrt{r})\right]
\nonumber\\
&&
-\frac{1}{2} \log \frac{1+\sqrt{1-r}}{1-\sqrt{1-r}}
.~~~\label{int-root}%
\end{eqnarray}
\end{subequations}
The integrals over $x_1$ of the infrared-divergent functions 
$\hat{\Gamma}_{\textrm{div}}^J(x_1) $ given in  
Eq.~(\ref{div-eval}) are 
\begin{subequations}
\begin{eqnarray}
\int_{\sqrt{r}}^{1}dx_1
\hat{\Gamma}_{\textrm{div}}^0(x_1)
 &=&
\frac{2(2+r)}{9}
     \left[ -\frac{1}{2\epsilon}
            -\log\frac{r}{8}+3\log\sqrt{1-r}
     \right]
\sqrt{1-r}
\nonumber\\
&&
-\frac{8+r}{27}
\sqrt{1-r}
-\frac{4+3r}{9}\log \frac{1+\sqrt{1-r}}{1-\sqrt{1-r}} 
,\\
\int_{\sqrt{r}}^{1}dx_1
\hat{\Gamma}_{\textrm{div}}^1(x_1)
&=&
\frac{2(2+r)}{9}
     \left[ -\frac{1}{2\epsilon}
            -\log\frac{r}{8}+3\log\sqrt{1-r}
     \right]
\sqrt{1-r}
\nonumber\\
&&
-\frac{2(4+5r)}{27}
\sqrt{1-r}
-\frac{8+3r}{18}\log \frac{1+\sqrt{1-r}}{1-\sqrt{1-r}} 
,\\
\int_{\sqrt{r}}^{1}dx_1
\hat{\Gamma}_{\textrm{div}}^2(x_1)
&=&
\frac{2(2+r)}{9}
     \left[ -\frac{1}{2\epsilon}
            -\log\frac{r}{8}+3\log\sqrt{1-r}
     \right]
\sqrt{1-r}
\nonumber\\
&&
-\frac{8(5+4r)}{135}
\sqrt{1-r}
-\frac{40+21r}{90}\log\ \frac{1+\sqrt{1-r}}{1-\sqrt{1-r}} .
\end{eqnarray}
\end{subequations}
The integrals over $x_1$ of the infrared-finite functions 
$\hat{\Gamma}_{\textrm{fin}}^J(x_1) $ given in  
Eq.~(\ref{fin-eval}) are 
\begin{subequations}
\label{gamma_fin_tot}%
\begin{eqnarray}
\int_{\sqrt{r}}^{1}dx_1
\hat{\Gamma}_{\textrm{fin}}^0(x_1) &=&
-\frac{8(2+r)}{9} \sqrt{1-r}
+\frac{3+r}{3}
  \log \frac{1+\sqrt{1 - r}}{1-\sqrt{1 - r}} 
,\\
\int_{\sqrt{r}}^{1}dx_1
\hat{\Gamma}_{\textrm{fin}}^1(x_1) &=&
-\frac{2+r}{9} \sqrt{1-r}
+\frac{r}{6}
  \log \frac{1+\sqrt{1 - r}}{1-\sqrt{1 - r}} 
,\\
\int_{\sqrt{r}}^{1}dx_1
\hat{\Gamma}_{\textrm{fin}}^2(x_1) &=&
-\frac{22+23r}{45} \sqrt{1-r}
+\frac{8+7r}{30}
  \log \frac{1+\sqrt{1 - r}}{1-\sqrt{1 - r}} 
.
\end{eqnarray}
\end{subequations}

\end{document}